\documentclass{aa}  

\usepackage{graphicx}
\usepackage{xcolor}
\usepackage{txfonts}
\begin{document}

\title{The VMC survey -- XLV. Proper motion of the outer LMC and the impact of the SMC\thanks{Based on observations made with VISTA at the La Silla Paranal Observatory under programme ID 179.B-2003.}}

\authorrunning{Schmidt et al.}
\titlerunning{Proper motion of the outer LMC}

\author{Thomas~Schmidt\inst{1,2},
        Maria-Rosa~L.~Cioni\inst{1},
        Florian~Niederhofer\inst{1},
        Kenji~Bekki\inst{3},
        Cameron~P.~M.~Bell\inst{1},
        Richard~de~Grijs\inst{4,5},
        Dalal~El~Youssoufi\inst{1,2}, 
        Valentin~D.~Ivanov\inst{6},
        Joana~M.~Oliveira\inst{7},
        Vincenzo~Ripepi\inst{8},
        Jacco~Th.~van~Loon\inst{7}
        }

\institute{Leibniz-Institut für Astrophysik Potsdam (AIP), An der Sternwarte 16, D-14482 Potsdam, Germany\\
           \email{tschmidt@aip.de}
           \and
          Institut f\"{u}r Physik und Astronomie, Universit\"{a}t Potsdam, Haus 28, Karl-Liebknecht-Str. 24/25, D-14476 Golm (Potsdam), Germany
          \and
          ICRAR, M468, The University of Western Australia, 35 Stirling Hwy, Crawley WA 6009, Australia
          \and
          Department of Physics and Astronomy, Macquarie University, Balaclava Road, Sydney, NSW 2109, Australia
          \and
          Research Centre for Astronomy, Astrophysics and Astrophotonics, Macquarie University, Balaclava Road, Sydney, NSW 2109, Australia
          \and
          European Southern Observatory, Karl-Schwarzschild-Str. 2, D-85748 Garching bei München, Germany
          \and
          Lennard-Jones Laboratories, School of Chemical and Physical Sciences, Keele University, ST5 5BG, UK
          \and
          INAF-Osservatorio Astronomico di Capodimonte, via Moiariello 16, I-80131 Naples, Italy
            }

   \date{Received 3 September 2021; accepted 20 January 2022}

 
  \abstract
   {The Large Magellanic Cloud (LMC) is the most luminous satellite galaxy of the Milky Way and owing to its companion, the Small Magellanic Cloud (SMC), represents an excellent laboratory to study the interaction of dwarf galaxies.}
   {The aim of this study is to investigate the kinematics of the outer regions of the LMC by using stellar proper motions to understand the impact of interactions, e.g. with the SMC about 250 Myr ago.}
   {We calculate proper motions using multi-epoch $K_\mathrm{s}$-band images from the VISTA survey of the Magellanic Clouds system (VMC). Observations span a time baseline of 2$-$5 yr. We combine the VMC data with data from the \textit{Gaia} early Data Release 3 and introduce a new method to distinguish between Magellanic and Milky Way stars based on a machine learning algorithm. This new technique enables a larger and cleaner sample selection of fainter sources as it reaches below the red clump of the LMC.}
   {We investigate the impact of the SMC on the rotational field of the LMC and find hints of stripped SMC debris. The south east region of the LMC shows a slow rotational speed compared to the overall rotation. $N$-body simulations suggest that this could be caused by a fraction of stripped SMC stars, located in that particular region, that move opposite to the expected rotation.}
   {}


\maketitle
\section{Introduction}
\label{ch:intro}

Proper motion studies help us to understand key aspects of galaxy formation in the Local Group. One of these aspects is identifying past merger events. The Milky Way (MW) for example experienced multiple merger events \citep[e.g.][]{helmi2018,Kruijssen2020} with at least one major (i.e. with a galaxy of comparable mass) and multiple minor (i.e. with dwarf galaxies with a mass ratio of less than 1:10) events. The most massive satellite \citep[$\sim$6$\times 10^{10}$ M$_\odot$;][]{Mucciarelli2017} to have merged with the MW in the recent past was the Sagittarius dwarf galaxy. Its remains are visible over large parts of the sky, owing to the tidal disruption by the MW. This event depicts currently a final stage of merging with mainly the core of the dwarf galaxy still remaining. Another very similar, but less evident example, is the merging of the Canis Major dwarf galaxy \citep[e.g.][]{Martin2004}. The next similar merging event will most likely be with the Large Magellanic Cloud \citep[in at least 7 Gyr; e.g.][]{Hashimoto2003} after multiple passages. The Large and Small Magellanic Clouds (LMC and SMC) are the most luminous and massive dwarf galaxies around the MW. They both are classified as irregular dwarf galaxies located at a distance of 50$-$60 kpc. They are in an early stage of merging with the MW, as they are probably completing their first orbit around it (e.g.~\citealp{Besla2007,vandermarel2014,Hammer2015,Patel2017}).

In addition to interacting with the MW, the Magellanic Clouds also interact with one another. The SMC is believed to have had at least two close interactions with the more massive LMC (e.g. \citealp{Besla2016}; \citealp{Pearson2018}). The Magellanic Clouds currently provide an excellent opportunity to study in detail the kinematics of resolved stellar populations in an interacting pair of galaxies.
The total mass of the LMC is still a matter of debate. More recent studies derived it to be larger than previously estimated, e.g.~$\sim$1.4$\times$10$^{11}$M$_\odot$ (\citealp{Erkal2019}). The mass of the SMC is also unclear owing to the tidal influence of the LMC. Their interaction alone stripped a significant amount of mass (gas and stars) from the SMC which formed the Magellanic Bridge and Stream (e.g. \citealp{gardiner1996}; \citealp{diaz2012}; \citealp{besla2013}). The atomic gas outflow (0.2$-$1 M$_\odot$ yr$^{-1}$) exceeds by one order of magnitude the rate of star formation within the galaxy (\citealp{mcclure-griffiths2018}). The density and velocity flow of stars towards the LMC along the Bridge is clearly traced in recent studies (e.g. \citealp{zivick2019}; \citealp{schmidt2020}; \citealp{luri2020}).

The internal kinematics of the SMC stars, which is completely dominated by the dynamical interactions, exhibits essentially no measurable rotation (e.g. \citealp{zivick2020}; Niederhofer et al. 2020) contrary to a pronounced rotation of the HI gas (\citealp{stanimirovic2004}; \citealp{diteodoro2019}). The stellar motion appears consistent with tidal stripping or stretching of the galaxy. The more massive LMC seems to be less affected by these events, but dynamical interactions likely caused the LMC bar to be located off centre and increased star formation (e.g. the 30 Doradus starburst region), whereas the outer regions show a clock-wise rotation (e.g. \citealp{olsen2011}; \citealp{vandermarel2014}; \citealp{helmi2018}). Just like the Sagittarius dwarf galaxy in the MW, we expect that material of the SMC would be distributed throughout the LMC.

In this study we derive proper motions from multi-epoch $K_\mathrm{s}$-band observations from the Visible and Infrared Survey Telescope for Astronomy (VISTA) survey of the Magellanic Clouds system (VMC; \citealp{cioni2011}) to investigate the influence of SMC stars on the LMC's internal kinematics. The proper motion measures two key components of the three dimensional (space) velocity which are of great importance to understand the kinematics of a nearly face-on LMC disc. 
We combine VMC data with data from the \textit{Gaia} early Data Release 3 (EDR3; \citealp{brown2020}) to remove the influence of MW stars. Both VMC and \textit{Gaia} EDR3 data provide continuous coverage across the LMC. In addition, we develop a technique to remove MW stars also at faint magnitudes which are not yet accounted for in the Bayesian inference code {\sc StarHorse} \citep{Anders2021} to estimate distances. Our study is focused on the outer regions of the LMC which are not affected by crowding in either the VISTA or \textit{Gaia} datasets.

The paper is structured as follows: Section~\ref{ch:Observations} describes the VMC observations used in this study. Section \ref{ch:Analysis} describes the steps of the analysis starting with the VMC proper motions and followed by the $Gaia$ sample selection. Then, we introduce a machine learning classification algorithm to derive Magellanic stellar memberships and to remove MW foreground stars. The resulting proper motion maps are presented in Section~\ref{LMC_kinematics}. Section~\ref{ch:Conclusions} concludes the paper.

\section{Observations}
\label{ch:Observations}

\begin{figure*}
   \centering
   \includegraphics[width=1.0\textwidth]{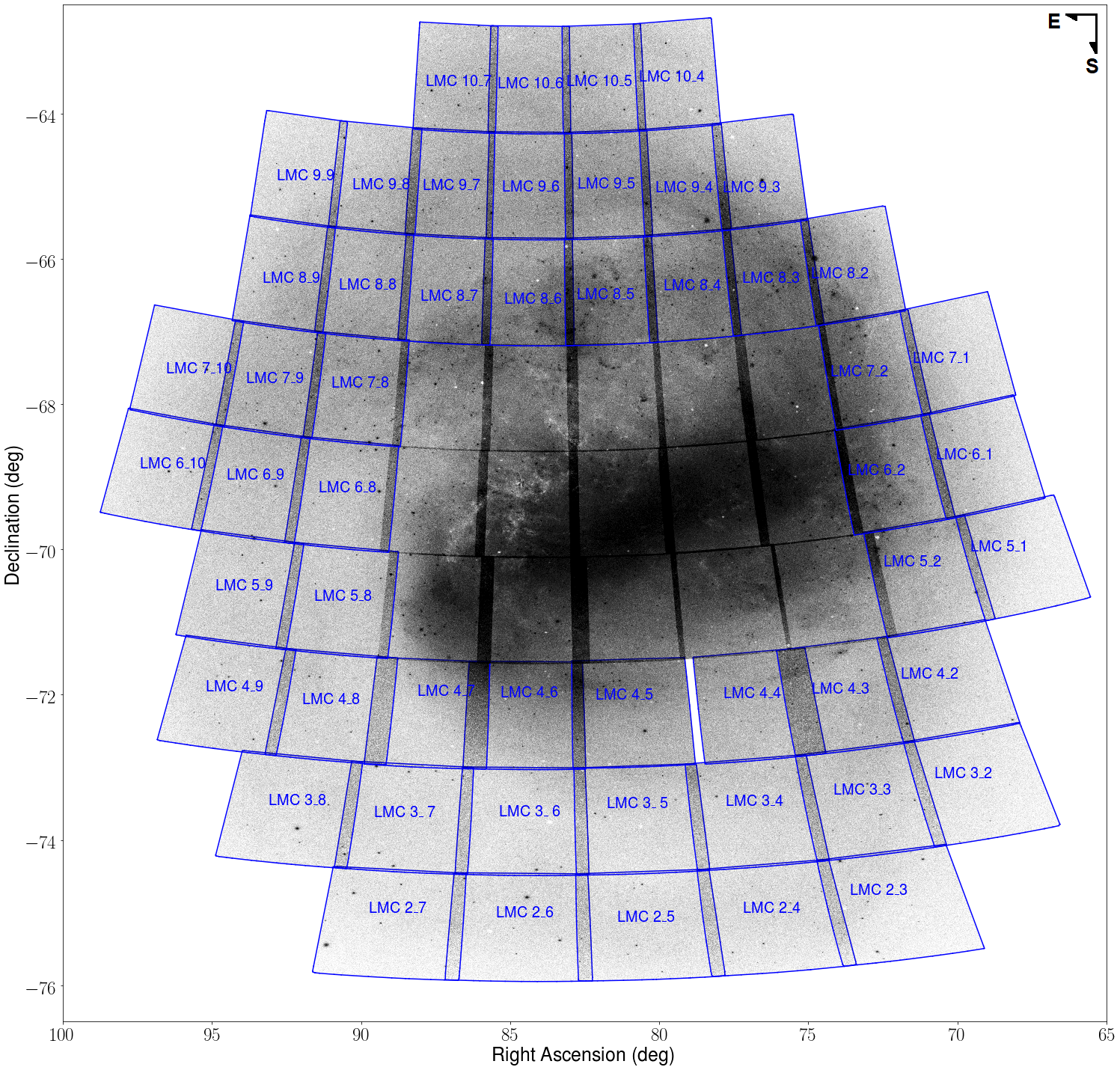}
   
   \caption{Distribution of VMC tiles across the LMC. The tiles used in this study are those in the outer regions of the galaxy and are indicated by their IDs. The image in the background shows the distribution of all VMC sources.}
    \label{LMC_tiles}%
    \end{figure*}

Data analysed in this study are taken from the VMC survey (Cioni et al. 2011). The VMC survey was completed in October 2018 and gathered multi-epoch near-infrared images in the $Y$, $J$, and $K_\mathrm{s}$ filters of 110 overlapping tiles across the Magellanic system: 68 covering the LMC, 27 the SMC, 13 the Bridge, and two the Stream components. Each tile covers an area of 1.77~deg$^{2}$ on the sky and about 1.5~deg$^{2}$ with multiple observations. In this study, we focus on the outer LMC which is covered by 53 tiles. The distribution of these tiles can be seen in Fig.~\ref{LMC_tiles}. The most central 15 tiles distributed along the bar of the LMC are left out of this study because they present a high density of stars and possible crowding issues; they are therefore addressed in a separate study (\citealp{niederhofer2021b}).

Observations were obtained with the VISTA Infrared Camera (VIRCAM) mounted on VISTA\footnote{http://www.vista.ac.uk} \citep{Sutherland2015} which is operated by the European Southern Observatory (ESO). VIRCAM is a near-infrared imaging camera composed of 16 VIRGO HgCdTe detectors. Each detector covers an area of 0.0372~deg$^2$ with an average pixel scale of 0.339$^{\prime\prime}$. The individual images from the 16 detectors form a VISTA pawprint that covers 0.6~deg$^2$, not including the gaps between the detectors. A mosaic of 6 pawprints was used to cover a contiguous area filling the gaps between individual detectors. This arrangement forms a VMC tile. The individual detector integration time (DIT) for a $K_{s}$-band exposure was 5~s. Taking 5 jitters and 15 repetitions into account, each point on the sky is observed by two different pixels, so the total exposure amounts to 750~s per tile and epoch. However, in a single pawprint each pixel is exposed on average for 375~s per tile. There are at least 11 epochs at $K_\mathrm{s}$ of this type (deep) and two epochs with half the exposure time (shallow). Exposure times in the $Y$ and $J$ bands as well as additional parameters of the survey are described in detail by \citet{cioni2011}. Images were processed using the VISTA Data Flow System pipeline \citep[VDFS v1.5,][]{Emerson2006} at the Cambridge Astronomy Survey Unit\footnote{http://casu.ast.cam.ac.uk} (CASU) and stored in the VISTA Science Archive\footnote{http://horus.roe.ac.uk/vsa} \citep[VSA,][]{Cross2012}. The catalogues provided by the VSA contain aperture photometry which is sufficient to study the relatively moderate stellar density (80.000 stars per deg$^{2}$ on average) across the outer regions of the LMC. Magnitudes have been calibrated as explained in \cite{gonzalez2018} and result in an accuracy of better than 0.02 mag in $YJK_\mathrm{s}$. The astrometric calibration of the VMC data is based on the Two Micron All Sky Survey (2MASS, \citealp{skrutskie2006}) and carries a systematic uncertainty of 10$-$20 mas due to World Coordinate System errors\footnote{http://casu.ast.cam.ac.uk/surveys-projects/vista/technical/astrometric-properties}. Those are systematic uncertainties in the calibration of each detector image obtained using 2MASS stars. They are mainly caused by atmospheric turbulence and atmospheric differential refraction. Table \ref{obs_table} provides details about the observations. It contains the tile identification, the central coordinates, the orientation, the number of epochs used, their time baseline, the FWHM, the airmass and the sensitivity\footnote{http://casu.ast.cam.ac.uk/surveys-projects/vista/technical/vista-sensitivity}, derived from sources with photometric uncertainties <0.1 mag. The average values of all good quality deep $K_\mathrm{s}$ epochs (11 for most tiles) were 0.93$\pm$0.10~arcsec (FWHM), 1.49$\pm$0.09 (Airmass) and 19.28$\pm$0.16~mag (Sensitivity).

\begin{table*}[] 
\setlength{\tabcolsep}{1pt}
    \caption{VMC $K_\mathrm{s}$-band observations of the LMC tiles.}
    \centering
    \begin{tabular}{lccrccccc}
    \hline\hline
    Tile & Right Ascension$^\mathrm{a}$ & Declination$^\mathrm{a}$ & Position angle$^\mathrm{b}$ & Epochs & Baseline & FWHM$^\mathrm{c}$ & Airmass$^\mathrm{c}$ & Sensitivity$^\mathrm{c,d}$ \\
     & (h:m:s) & ($^\circ$:$^\prime$:$^{\prime\prime}$) & (deg) & & (day) & (arcsec) & & (mag) \\
    \hline
LMC 2\_3 & 04:48:04.752 & $-$74:54:11.880 & $-$101.226 & 11 & 1510 & 0.99$\pm$0.15 & 1.60$\pm$0.04 & 19.26\\
LMC 2\_4 & 05:04:42.696 & $-$75:04:45.120	& $-$97.320 & 11 & 1134 & 0.97$\pm$0.11 & 1.60$\pm$0.04 & 19.31 \\
LMC 2\_5 & 05:21:38.664 & $-$75:10:50.160	& $-$93.338 & 11 & 1532 & 0.98$\pm$0.09 & 1.59$\pm$0.02 & 19.31 \\
LMC 2\_6 & 05:38:43.056 & $-$75:12:21.240 & $-$89.321 & 11 & 1606 & 0.90$\pm$0.08 & 1.60$\pm$0.02 & 19.23 \\
LMC 2\_7 & 05:55:45.720 & $-$75:09:17.280 & $-$85.312 & 11 & 807 & 0.95$\pm$0.10 & 1.60$\pm$0.02 & 19.32 \\
LMC 3\_2 & 04:37:05.256 & $-$73:14:30.120 & $-$103.773 & 11 & 1531 & 0.97$\pm$0.08 & 1.57$\pm$0.06 & 19.31 \\
LMC 3\_3 & 04:51:59.640 & $-$73:28:09.120 & $-$100.284 & 12 & 1515 & 0.93$\pm$0.09 & 1.56$\pm$0.04 & 19.32 \\
LMC 3\_4 & 05:07:14.472 & $-$73:37:49.800 & $-$96.710 & 11 & 1798 & 0.93$\pm$0.09 & 1.60$\pm$0.07 & 19.29\\
LMC 3\_5 & 05:22:43.056 & $-$73:43:25.320 & $-$93.079 & 11 & 693 & 0.94$\pm$0.09 & 1.60$\pm$0.04 & 19.40 \\
LMC 3\_6 & 05:38:18.096 & $-$73:44:51.000 & $-$89.421 & 11 & 1593 & 0.94$\pm$0.07 & 1.57$\pm$0.05 & 19.16 \\
LMC 3\_7 & 05:53:51.912 & $-$73:42:05.760 & $-$85.767 & 11 & 1171 & 0.98$\pm$0.11 & 1.55$\pm$0.02 & 19.24 \\
LMC 3\_8 & 06:09:16.920 & $-$73:35:12.120 & $-$82.151 & 11 & 1522 & 0.91$\pm$0.13 & 1.54$\pm$0.02 & 19.15 \\
LMC 4\_2 & 04:41:30.768 & $-$71:49:16.320 & $-$102.717 & 11 & 1103 & 0.94$\pm$0.10 & 1.54$\pm$0.08 & 19.44 \\
LMC 4\_3 & 04:55:19.512 & $-$72:01:53.400 & $-$99.488 & 11 & 1088 & 0.92$\pm$0.10 & 1.59$\pm$0.09 & 19.41 \\
LMC 4\_4 & 05:09:32.496 & $-$72:10:16.680 & $-$96.277 & 11 & 1066 & 0.94$\pm$0.09 & 1.52$\pm$0.04 & 19.27 \\
LMC 4\_5 & 05:23:46.560 & $-$72:15:21.960 & $-$92.478 & 11 & 740 & 0.86$\pm$0.08 & 1.52$\pm$0.04 & 19.37 \\
LMC 4\_6 & 05:38:00.408 & $-$72:17:20.040 & $-$89.491 & 11 & 992 & 0.92$\pm$0.09 & 1.52$\pm$0.04 & 19.29 \\
LMC 4\_7 & 05:50:50.496 & $-$72:15:39.960 & $-$86.814 & 11 & 732 & 0.85$\pm$0.06 & 1.52$\pm$0.03 & 19.37 \\
LMC 4\_8 & 06:03:40.872 & $-$72:10:06.240 & $-$83.475 & 11 & 990 & 0.99$\pm$0.09 & 1.53$\pm$0.04 & 19.33 \\
LMC 4\_9 & 06:17:43.560 & $-$72:00:48.240 & $-$80.188 & 11 & 1505 & 0.91$\pm$0.10 & 1.53$\pm$0.07 & 19.19 \\
LMC 5\_1 & 04:31:28.032 & $-$70:06:57.600 & $-$105.047 & 11 & 1010 & 0.99$\pm$0.08 & 1.52$\pm$0.09 & 19.33 \\
LMC 5\_2 & 04:44:01.728 & $-$70:22:21.000 & $-$102.118 & 11 & 1023 & 0.83$\pm$0.04 & 1.52$\pm$0.08 & 19.40 \\
LMC 5\_8 & 06:02:56.232 & $-$70:42:25.920 & $-$83.655 & 11 & 1564 & 0.96$\pm$0.12 & 1.47$\pm$0.04 & 19.21 \\
LMC 5\_9 & 06:15:59.112 & $-$70:33:27.360 & $-$80.604 & 11 & 1644 & 0.91$\pm$0.09 & 1.52$\pm$0.05 & 19.31 \\
LMC 6\_1 & 04:36:49.488 & $-$68:43:50.880 & $-$103.794 & 11 & 1037 & 0.95$\pm$0.10 & 1.51$\pm$0.09 & 19.23 \\
LMC 6\_2 & 04:48:39.072 & $-$68:57:56.520 & $-$101.035 & 11 & 970 & 0.86$\pm$0.07 & 1.48$\pm$0.08 & 19.32 \\
LMC 6\_8 & 06:02:21.984 & $-$69:14:42.360 & $-$83.790 & 11 & 824 & 0.88$\pm$0.09 & 1.48$\pm$0.04 & 19.42 \\
LMC 6\_9 & 06:14:32.832 & $-$69:05:59.640 & $-$80.943 & 11 & 1702 & 0.91$\pm$0.08 & 1.52$\pm$0.09 & 19.29 \\
LMC 6\_10 & 06:26:32.280 & $-$68:54:05.760 & $-$78.142 & 11 & 1514 & 0.94$\pm$0.11 & 1.49$\pm$0.11 & 19.23 \\
LMC 7\_1 & 04:40:09.167 & $-$67:18:19.800 & $-$103.023 & 11 & 1030 & 1.00$\pm$0.12 & 1.50$\pm$0.11 & 19.33 \\
LMC 7\_2 & 04:51:17.832 & $-$67:31:39.000 & $-$100.421 & 11 & 740 & 0.88$\pm$0.03 & 1.43$\pm$0.09 & 19.35 \\
LMC 7\_8 & 06:00:27.696 & $-$67:47:48.120 & $-$84.234 & 11 & 1493 & 0.96$\pm$0.07 & 1.46$\pm$0.09 & 19.20 \\
LMC 7\_9 & 06:11:54.384 & $-$67:39:41.400 & $-$81.557 & 11 & 1706 & 0.93$\pm$0.07 & 1.49$\pm$0.10 & 19.21 \\
LMC 7\_10 & 06:23:11.736 & $-$67:28:34.320 & $-$78.918 & 12 & 1515 & 0.97$\pm$0.15 & 1.45$\pm$0.07 & 19.18 \\
LMC 8\_2 & 04:54:11.568 & $-$66:05:47.760 & $-$99.755 & 11 & 1051 & 0.99$\pm$0.11 & 1.41$\pm$0.10 & 19.31 \\
LMC 8\_3 & 05:04:53.952 & $-$66:15:29.880 & $-$97.249 & 11 & 705 & 0.96$\pm$0.10 & 1.38$\pm$0.03 & 19.43 \\
LMC 8\_4 & 05:15:43.464 & $-$66:22:19.920 & $-$94.713 & 11 & 1032 & 0.96$\pm$0.11 & 1.43$\pm$0.09 & 19.26 \\
LMC 8\_5 & 05:26:37.152 & $-$66:26:15.720 & $-$92.160 & 11 & 693 & 0.90$\pm$0.07 & 1.41$\pm$0.07 & 19.35 \\
LMC 8\_6 & 05:37:34.104 & $-$66:27:15.840 & $-$89.593 & 11 & 1103 & 0.95$\pm$0.08 & 1.42$\pm$0.08 & 19.24 \\
LMC 8\_7 & 05:48:30.120 & $-$66:25:19.920 & $-$87.030 & 11 & 1506 & 0.98$\pm$0.10 & 1.46$\pm$0.10 & 19.16 \\
LMC 8\_8 & 05:59:23.136 & $-$66:20:28.680 & $-$84.480 & 11 & 377 & 0.91$\pm$0.10 & 1.38$\pm$0.05 & 19.41 \\ 
LMC 8\_9 & 06:10:10.632 & $-$66:12:43.560 & $-$81.953 & 11 & 1515 & 0.89$\pm$0.17 & 1.42$\pm$0.08 & 19.16 \\
LMC 9\_3 & 05:06:40.632 & $-$64:48:40.320 & $-$96.844 & 11 & 1076 & 0.98$\pm$0.10 & 1.40$\pm$0.09 & 19.35 \\ 
LMC 9\_4 & 05:16:39.792 & $-$64:54:59.760 & $-$94.501 & 11 & 1335 & 0.96$\pm$0.09 & 1.45$\pm$0.08 & 19.32 \\
LMC 9\_5 & 05:26:58.512 & $-$64:58:45.840 & $-$92.080 & 11 & 1340 & 0.92$\pm$0.09 & 1.44$\pm$0.11 & 19.25 \\
LMC 9\_6 & 05:37:19.104 & $-$64:59:45.240 & $-$89.651 & 11 & 1620 & 0.93$\pm$0.13 & 1.44$\pm$0.10 & 19.20 \\
LMC 9\_7 & 05:47:55.128 & $-$64:57:52.920 & $-$87.162 & 11 & 1031 & 0.93$\pm$0.11 & 1.38$\pm$0.06 & 19.48 \\
LMC 9\_8 & 05:57:57.168 & $-$64:53:24.360 & $-$84.807 & 11 & 1689 & 0.89$\pm$0.08 & 1.48$\pm$0.10 & 19.30 \\
LMC 9\_9 & 06:08:10.343 & $-$64:46:05.880 & $-$82.409 & 11 & 1477 & 0.93$\pm$0.09 & 1.42$\pm$0.07 & 19.26 \\
LMC 10\_4 & 05:17:46.656 & $-$63:27:46.440 & $-$94.249 & 11 & 1688 & 0.90$\pm$0.08 & 1.46$\pm$0.12 & 19.27 \\
LMC 10\_5 & 05:27:33.096 & $-$63:31:19.200 & $-$91.949 & 11 & 972 & 0.96$\pm$0.07 & 1.35$\pm$0.07 & 19.38 \\
LMC 10\_6 & 05:37:22.848 & $-$63:32:13.560 & $-$89.636 & 11 & 1444 & 0.92$\pm$0.11 & 1.41$\pm$0.08 & 19.16 \\
LMC 10\_7 & 05:47:11.424 & $-$63:30:29.520 & $-$87.327 & 12 & 1129 & 0.95$\pm$0.08 & 1.39$\pm$0.09 & 19.30 \\
      \hline
    \end{tabular}
    
    $^\mathrm{(a)}$ Coordinates of VMC tile centres.
    $^\mathrm{(b)}$ Orientation of the VMC tiles, defined to increase from North to East.
    $^\mathrm{(c)}$ Average of all used epochs.
    $^\mathrm{(d)}$ For sources with photometric uncertainty <0.1 mag.
\label{obs_table}
\end{table*}

\section{Analysis}
\label{ch:Analysis}

\subsection{VMC proper motions} \label{VMCpms}
The selection of the sample of VMC sources for which to derive the proper motions, the creation of the astrometric reference frame and the calculation of the proper motions follow the same steps as described in \cite{schmidt2020}. Some of the steps are however repeated here for clarity.

\subsubsection{Sample selection} \label{VMCsample}

The VMC source catalogues for each tile were obtained from the VSA using a freeform SQL query. We extracted equatorial coordinates (Right Ascension, Declination) in J2000, source-type classifications ($\textit{mergedClass}$), magnitudes ($J$ and $K_\mathrm{s}$), the corresponding uncertainties and quality extraction flags (\textit{ppErrBits}) for each source. VMC tiles, pawprints and sources in the VSA are identified by their identification numbers: tiles by a unique \textit{framesetID}, individual pawprints by a unique \textit{multiframeID} and individual sources by a unique \textit{sourceID}. The source-type classification flags were used to distinguish between stars and galaxies while quality extraction flags were used to remove low-quality detections, by limiting the \textit{ppErrBits} to 16. This selection criterion removes VMC sources with systematic uncertainties affecting the photometric calibration. The VSA \textit{vmcdetection} table contains data of the individual pawprints originating from stacked images. Source catalogues based on individual epoch observations were obtained by cross-matching the list of sources with those in the \textit{vmcdetection} tables retaining all matches within 0.5$^{\prime\prime}$. The resulting catalogue contains the mean Modified Julian Day (MJD) of the observation, the detector number (\textit{extNum}), the pixel coordinates ($x$, $y$) on each detector and the corresponding positional uncertainties.
We split the VMC epoch catalogues into 96 parts (for 6 pawprints $\times$ 16 detectors) per epoch and tile, respectively. 
Distinct epochs were then selected based on their \textit{multiframeIDs}. All sources with the same \textit{multiframeID} are part of the same pawprint observed across the 16 detectors. Undesired \textit{multiframeIDs} such as those associated with observations from overlapping tiles (where sources would be detected in different detectors), observations obtained under poor sky conditions, and detections at wavelengths other than $K_\mathrm{s}$ were removed. Every catalogue was then divided into two parts. One contains only sources classified as galaxies ($\textit{mergedClass}$=$1$) and the other contains only stars ($\textit{mergedClass}$=$-$1). We rejected all other source-type classifications (e.g.~noise, probable stars and probable galaxies). In a small number of cases two sources in the \textit{vmcdetection} tables, with the same \textit{multiframeIDs}, were matched to one \textit{sourceID} from the \textit{vmcsource} catalogue. This duplication was caused by the matching algorithm when two sources were sufficiently close together in the detection catalogue while one of the sources was missing in the \textit{vmcsource} catalogue. The nearest source was selected in this case.

\subsubsection{Deriving the proper motions}

To calculate consistent proper motions, each observation of a given source has to be in the same astrometric reference frame. The reference frames for each VMC tile in this study were created by choosing the epoch with the best observing conditions from each set of observations of a given tile. This corresponds to the epoch with the largest number of extracted sources and the smallest FWHM. The reference frames were constructed using background galaxies. The number of background galaxies in the VMC survey is quite large (a few hundred per detector) and in the outer regions of the LMC there are >200 per detector with >300 in less crowded regions. This is sufficient to produce proper motions as accurate as those that can be derived from a reference system made of VMC stars within the same area (\citealp{niederhofer2018a}).
The median rms values of matching the epochs using background galaxies was 0.23 pixels and all matches had a value smaller than 0.28 pixels. These residuals affect the median proper motion of the background galaxies, resulting in a moving reference frame. To correct the co-moving reference frame, we set the $\sigma$-clipped relative median proper motion of the galaxies for each detector to zero. We checked for possible systematic effects, such as unevenly distributed samples, an influence of uncertainties in individual coordinates and the size of the matching samples. None of these seemed to have a significant influence on the results.

%
\begin{table}[]
\caption{Size of stellar samples}
\label{catalog_sizes}
\begin{tabular}{l r}        
\hline\hline                 
Catalogue & Number of stars\\    
\hline                        
VMC & 16,210,461 \\
\textit{Gaia}~eDR3 & {7,265,212} \\
StarHorse & {796,875} \\
LMC members (StarHorse) & {636,437}\\
LMC members (SVM) & {2,629,456}\\
\hline                                   
\end{tabular}
\end{table}

After creating the reference frame every corresponding epoch catalogue was transformed into it using IRAF  \citep{Tody1993} tasks  \textit{xyxymatch}, \textit{geomap}, and \textit{geoxytran} and then joined with the reference epoch catalogue of the same pointing and detector. The proper motions of individual stars were calculated by using a linear least-squares fit for the \textit{x} and \textit{y} coordinates separately, and the corresponding MJD with respect to a reference frame defined by background galaxies. Each fit contained typically 11 data points spanning an average time baseline of 1230 days (see Table \ref{obs_table} for the time baseline for each tile). Calculations were performed on a detector-by-detector basis for each of the 16 detectors and 6 pawprints of each tile. The slopes of these fits are the proper motions of individual stars for the two components in units of pixels per day. The conversion to mas yr$^{-1}$ was done using the World Coordinate System information from the FITS headers of the detector images at the reference epoch (see \citealt{cioni2016} for details).

When calculating the median proper motions of a selection of stars, we removed outliers using a 3$\sigma$ clipping technique where $\sigma$ was calculated using the median absolute deviation (MAD). The statistical error was calculated as the MAD divided by the square root of the numbers of stars. The $\sigma$-clipping was repeated until no additional sources were removed.
We checked the proper motions for any trends with detector number, position on the detectors and $J-K_\mathrm{s}$ colours and found nothing significant influencing our results. Finally, we derived the near-infrared proper motion of 16,210,461 stellar VMC sources in the outer regions of the LMC (Table~\ref{catalog_sizes}).

\subsection{Gaia EDR3 sample selection}\label{gaia_sample_selection}
$Gaia$~EDR3 data were acquired through the $Gaia$@AIP database\footnote{https://gaia.aip.de} which provides a cross-match with the VMC source catalogue using a tolerance radius of 1$^{\prime\prime}$ and taking the $Gaia$~EDR3 proper motions into account. The sky coordinates for sources in the VMC catalogue refer to the epoch J2000 whereas in the \textit{Gaia} EDR3 catalogue they refer to J2015.5; both datasets make use of the International Coordinate Reference System. We remove stars that could be problematic sources such as astrometric binaries, (partially) resolved binaries or multiple stars blended together using the re-normalised unit weight error (RUWE). As described in \cite{schmidt2020} and recommended by the $Gaia$ data processing and analysis consortium (DPAC), we select sources with $RUWE< 1.40$. A check of the residuals of VMC vs. Gaia proper motions shows that there are no significant correlations as a function of magnitude, colour or sky position (see Fig. \ref{comparison_VMC_Gaia}). VMC proper motions may be unreliable for individual sources, but binning a large number of them (at least a few hundred, depending on the level of contamination between stellar populations, e.g. between LMC and MW stars) shows a good agreement with the \textit{Gaia} values \citep{schmidt2020,niederhofer2021}. Figure \ref{Dispersion_Gaia_m_VMC} shows that the dispersion (calculated as maximum minus minimum) of the difference between $Gaia$ EDR3 and VMC proper motions decreases with increasing number of stars. Here, no distinction has been made between MW and LMC stars or different LMC stellar populations. All Voronoi bins of the VMC measurements (Fig.~\ref{LMC_rotaion_VMC}) agree to better then 0.05 mas yr$^{-1}$ because the smallest bin contains at least 1600 stars.

A subset of the $Gaia$~EDR3 catalogue, comprising stars with {$G<$18.5 mag}, was selected and combined with the newest {\sc StarHorse} distance estimates  (\citealp{Anders2021}) using the unique $Gaia$~sourceID, that both catalogues share. These sources form the basis of our training sample and will be further discussed in Section~\ref{SVMclassification}.
The VMC$-Gaia$ sample, including only sources with a measured parallax, contains less than half of the stellar sources in the VMC sample, in total 7,265,212 unique stellar sources, whereas the {\sc StarHorse} sample contains only 796,875 sources, due to the magnitude limitation. The sample sizes of the catalogues used in this study are shown in Table~\ref{catalog_sizes}.

\subsection{Distinguishing Magellanic Cloud from Milky Way stars}\label{Distinguishing_LMC_MW}
Our previous studies based on the VMC data (\citealp{niederhofer2018a,niederhofer2021}; \citealp{Schmidt2018}) emphasised the importance of an efficient foreground removal. Although \textit{Gaia} EDR3 delivers excellent parallax measurements, they still have significant uncertainties for objects at the
distance of the Magellanic Clouds. The same is true for $Gaia$ and VMC proper motions. Both parallax and proper motion measurements also become more challenging for fainter sources and therefore, for the vast majority of stars in the Magellanic Clouds. Simple cuts in parallax and proper motion are only sufficient to remove MW stars among the brightest stars, such as upper red giant branch and bright main sequence stars as shown in e.g.~\cite{Vasiliev2018}. 
The uncertainties in parallax and proper motions are significant enough that a large number of MW foreground stars are indistinguishable from the majority of the Magellanic Clouds stars based on parallaxes and proper motions alone. The nature of the overlapping functional distributions leads to an uneven distribution of MW foreground stars within the proper motion spread of the Magellanic Clouds stellar population as shown in \citet{schmidt2020} for the part of the LMC close to the Magellanic Bridge and \citet{helmi2018} for other dwarf galaxies (e.g. Sagittarius). This uneven distribution will always nudge every median measurement towards the MW proper motion value. While this effect is reducible by removing the majority of the foreground stars with other methods (e.g.~using colours and magnitudes) it will persist as long as there are MW stars in the sample.

More sophisticated methods are needed to significantly increase the sample size and enable better spatial resolutions to study the LMC kinematics in detail, while also efficiently removing the influence of the MW foreground. Using the distance estimates provided by the Bayesian inference in the {\sc StarHorse} code it is possible to obtain a clean sample of Magellanic Cloud stars, as shown in \cite{schmidt2020}. As described therein, this approach is however limited to stars with magnitudes $G<$18~mag in the previous {\sc StarHorse} version (\citealp{Queiroz2018}) and $G<$18.5~mag in the newest version (\citealp{Anders2021}). Our new approach bypasses the current magnitude limitation of the {\sc StarHorse} distances and enables a more detailed view of the outer LMC regions.

\subsubsection{Machine learning with SVM} \label{Machine learning}

We use a machine learning classification algorithm called support vector machines (SVM).
SVMs are binary large-margin classifiers and a supervised learning algorithm that evolved based on \citet{Boser1992} and \citet{Vapnik1995}. They allow to transfer classification information from a training set onto unseen data points, by mapping the training data with a non-linear transformation to a higher dimensional space to find a separating surface between the two classes. 
We use an approach similar to a galaxy cluster membership classification (\citealp{lopes2020}) to recover a membership probability from VMC and \textit{Gaia}~EDR3 data in order to remove efficiently MW foreground sources. Our method is based on a combination of astrometric and photometric properties (13 dimensions). They are parallax ($\omega$), parallax error ($\omega/\sigma_\omega$), the two components of proper motion ($\mu_\alpha cos\delta$, $\mu_\delta$) from \textit{Gaia} and VMC, and multiple magnitudes and colours ($J$, $K_{s}$, $J-K_{s}$, $G$, $G_{RP}$, $G_{BP}$ and $G_{RP-BP}$). 
Compared to the Bayesian inference method used in the {\sc StarHorse} code, SVM is computationally fast since the operation time is independent of the number of sources. As discussed in Section \ref{tests}, the SVM has been found to be good enough for the purpose of the paper.  In the future, radial velocity and element abundances for large samples of stars across the LMC will be provided by the 4-metre Multi-Object Survey Telescope (4MOST; \citealp{dejong2019,4mostCioni2019}) opening up new possibilities for a more precise membership classification using more parameters as well as a variety of classification methods (e.g. random forest regression, neural network).

\subsubsection{SVM classification}
\label{SVMclassification}
   \begin{figure}
   \centering
   \includegraphics[width=\hsize]{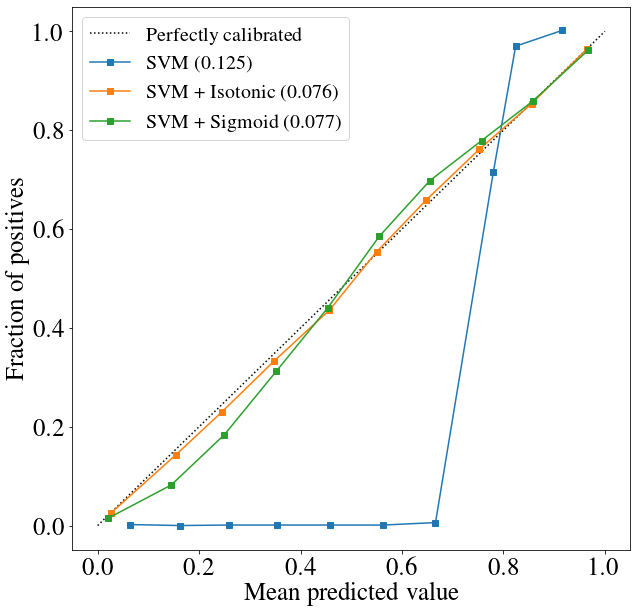}
      \caption{Reliability curves of the predicted values for the three different kernels used in the classification. The reliability curves compare the mean predicted value of a random subsample to the fraction of actual positives in the sample. The numbers between parentheses refer to the Brier score (see Sect.~\ref{Distinguishing_LMC_MW}).
              }
         \label{calibration_plot1}
   \end{figure}
   \begin{figure}
   \centering
   \includegraphics[width=\hsize]{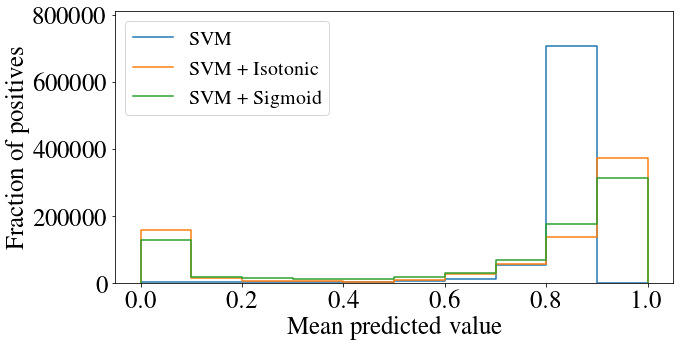}
      \caption{Histogram of the predicted values shown in Fig.~\ref{calibration_plot1}.
              }
         \label{calibration_plot2}
   \end{figure}

A large training set with reliable MW/LMC separation is hard to obtain and the combination with {\sc StarHorse}, with nearly 800,000 sources appears sufficient, leaving aside much smaller datasets, based on spectroscopic observations or special types of objects, that would not be appropriate for the problem in hand. The classification results for the validation sample represent a viable case for the overall classification, although the training sample is not fully representative as described below. 

Depending on the quality of the data and their uncertainties, it can occur that there is no hyperplane able to separate the two classes of objects (MW and LMC stars) in the training sample. In this case, the algorithm would not converge. To prevent this from happening, we decided to revise the $d=$20~kpc criterion used in \cite{schmidt2020} to separate between MW and Magellanic Bridge stars. The newest {\sc StarHorse} data provide a significantly better separation between MW and Magellanic Clouds stars, compared to the previous version, even though there is still an issue with underestimated distances, as shown in \citet{Anders2021}, in the form of an over-density between the MW and the LMC or other satellites. In the previous {\sc StarHorse} version many Magellanic Clouds sources ended up at significantly shorter or longer distances away from the bulk of the sources belonging to the galaxies. Therefore, stars of the LMC and most of the other MW satellites formed elongated cigar-like structures in all-sky density maps. By adding extra priors for the  satellites this issue was significantly reduced, especially in the outer regions of the Magellanic Clouds. Any remaining overdensity appears associated only with the central regions.

We classified stars with {\sc StarHorse} distance estimates smaller than 20~kpc as MW foreground stars and stars further than 20~kpc as LMC stars. These two samples define the base of the training and validation samples of the SVM classifier. Also, contrary to the previous {\sc StarHorse} code the updated version no longer systematically underestimates the distance of blue LMC stars, based on their $G_{BP}-G_{RP}$ colours. In the previous version those sources occupied the 10$-$40~kpc range where there was a significant overlap with the MW foreground population. This overlap blurred the separation between the two classes of objects and including those sources in the training sets reduced the predictive power of the SVM classification. The new {\sc StarHorse} version based on $Gaia$~EDR3 provides a significantly better training sample with respect to the LMC blue stars.

The VMC and \textit{Gaia}~EDR3 cross-matched catalogue contains 796,875 sources with {\sc StarHorse} distances. Of them, 553,235 can be clearly classified as either MW (160,438) or LMC stars (636,437), see Table \ref{catalog_sizes}. The actual training sample corresponds to a random fraction of 0.7 from both the MW and LMC. A fraction of 0.3 of the stars is used for validation. We used the python SVM package in the sklearn  library\footnote{https://scikit-learn.org/stable/modules/svm.html}. For a better classification and a higher flexibility of the hyperplane we tested multiple SVM kernels (e.g. Isotonic and Sigmoid) with a maximum of 100,000 iterations. Both the Isotonic and Sigmoid kernels enable more flexibility in the determination of the hyperplane that divides the training sets. This hyperplane is used to classify other sources. We found that the Isotonic calibration of the classifier delivered the highest prediction power with a Brier coefficient of 0.076 and a recall rate of 0.985. The Brier score describes the mean squared error of the prediction and ranges from 0 to 1, with 0 being a perfect calibration. The Isotonic regression provides high flexibility, but it is prone to overfitting \citep{Menon2012}. Our tests show that the training sample has sufficient data to avoid this, since the Isotonic regression outperforms the Sigmoid regression \citep{Niculescu-Mizil2005}.
 
The reliability curves of the standard (SVC), Isotonic and Sigmoid regressions are shown in Fig.~\ref{calibration_plot1}. The Isotonic calibration offers a nearly perfect calibration closely followed by the Sigmoid calibration, while the SVC is significantly biased towards the LMC. This is visible as a peak in the mean predicted value at 0.8 in the histogram (Fig.~\ref{calibration_plot2}). 
This indicates an increase of falsely classified sources which illustrates two important facts. First, that the majority of the falsely classified sources are classified as MW stars, this could be a direct effect of the bias introduced by {\sc StarHorse} underestimating the distances of LMC stars as mentioned before, which is carried over by residuals in the training sample. However, we noticed a significant improvement with respect to the previous {\sc StarHorse} version. The classifier became stricter as it classified more LMC stars as foreground stars than in the previous version. This could be due to the uncertainties in the $Gaia$~EDR3 measurements, many of the faint stars have too large uncertainties to be recognised as LMC stars.
Second, it proves that there is no simple solution possible in all the provided data columns. The simple SVM kernel does not find a clear separation between both classes as it classifies the majority of the foreground stars as LMC stars and therefore stricter selections will have a greater impact on the LMC sample size than on the fraction of MW foreground stars.

After training the classifier by variations of the hyperplane through random samples of the training set to increase the separation between the two classes (LMC and MW), the classifier was then used on the VMC--$Gaia$~EDR3 extended catalogue, including faint stars that do not have {\sc StarHorse} distances.
The SVM classification, using the Isotonic calibration, 
results in 2,629,456 LMC members (Table \ref{catalog_sizes}). 

   \begin{figure}
   \centering
   \includegraphics[width=\hsize]{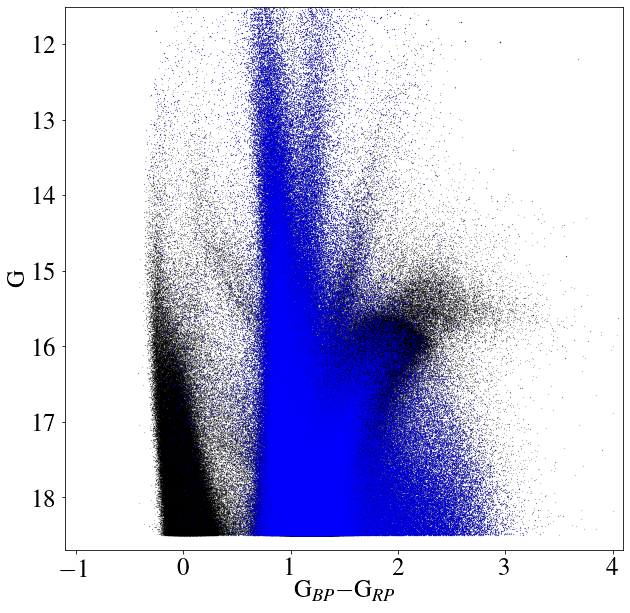}
   \includegraphics[width=\hsize]{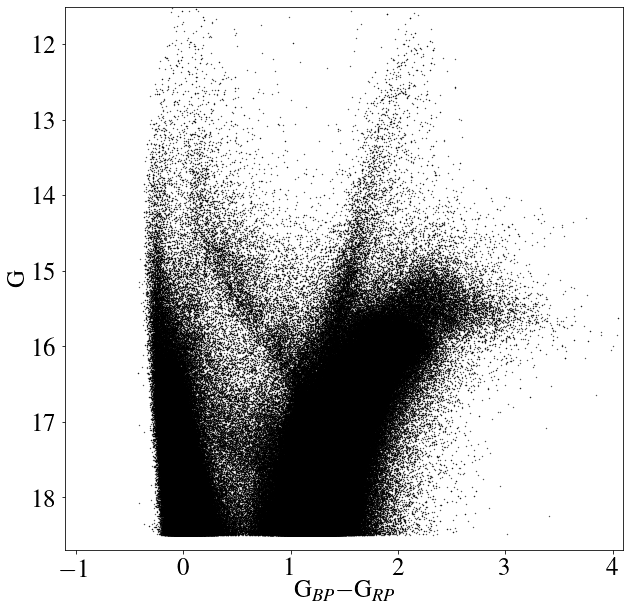}
      \caption{Top panel: Optical CMDs of sources in the LMC outer regions with VMC and \textit{Gaia} EDR3 photometry as well as {\sc StarHorse} distances (black) and the MW stars from our training sample having {\sc StarHorse} distance estimates $d<$10~kpc (blue). Bottom panel: Likely Magellanic stars with {\sc StarHorse} distance estimates $d>$20~kpc (based on \citealt{schmidt2020}). 
      }
         \label{CMD_MW_removal}
   \end{figure}

A critical factor in machine learning is the choice of the training sample. In \cite{schmidt2020}, we showed that {\sc StarHorse} distance estimates could be used to remove a significant fraction of MW foreground stars. Figure \ref{CMD_MW_removal} shows a comparison between the colour$-$magnitude diagram (CMD) of the VMC sources with a  \textit{Gaia} EDR3 counterpart and {\sc StarHorse} distances for LMC stars ($d>$20~kpc) as used in \cite{schmidt2020} and the majority of MW stars ($d<$10~kpc, blue). A selection based on {\sc StarHorse} distances is limited to stars with $G<$18.5 mag. The LMC stars in the {\sc StarHorse} sample are in fact only bright main sequence and giant stars, which excludes a large number of  faint Magellanic Cloud stars, such as the red clump stars (see Fig. \ref{cmd_populations} for the location of specific stellar populations). At faint magnitudes there is also a larger number of MW foreground stars.

\begin{figure}
   \centering
   \includegraphics[width=\hsize]{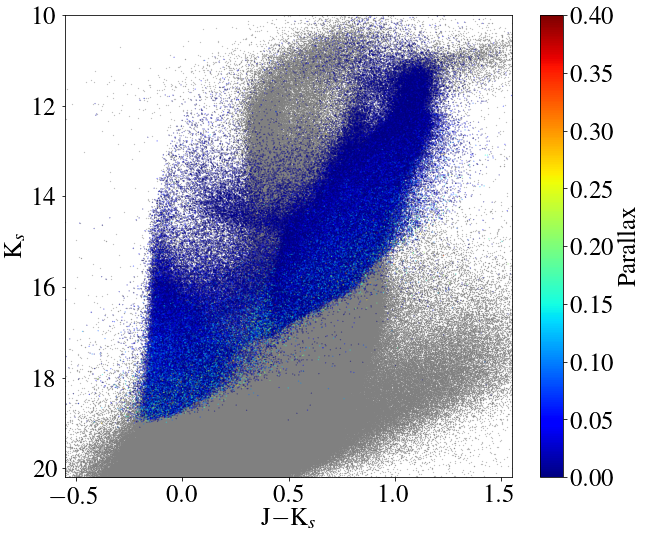}
   \includegraphics[width=\hsize]{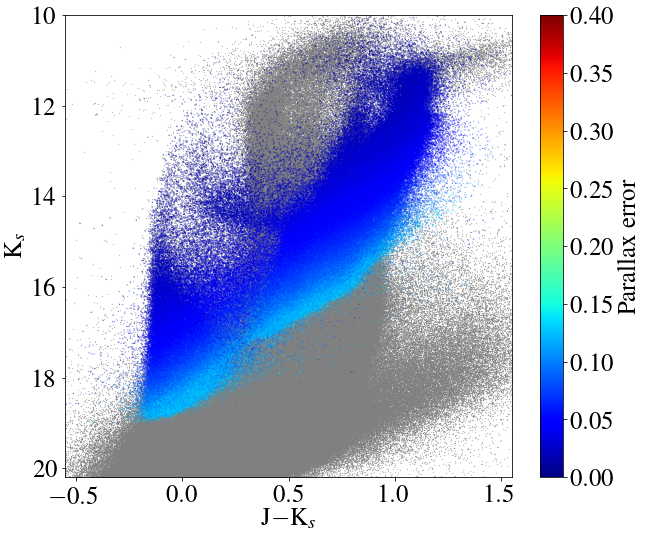}
      \caption{Near-infrared CMD of sources in the training sample colour coded by parallax (top) and parallax error (bottom) superimposed on the VMC and $Gaia$~EDR3 cross-match catalogue (grey).
              }
         \label{ML_LMC_aux}
   \end{figure}

\begin{figure}
   \centering
   \includegraphics[width=\hsize]{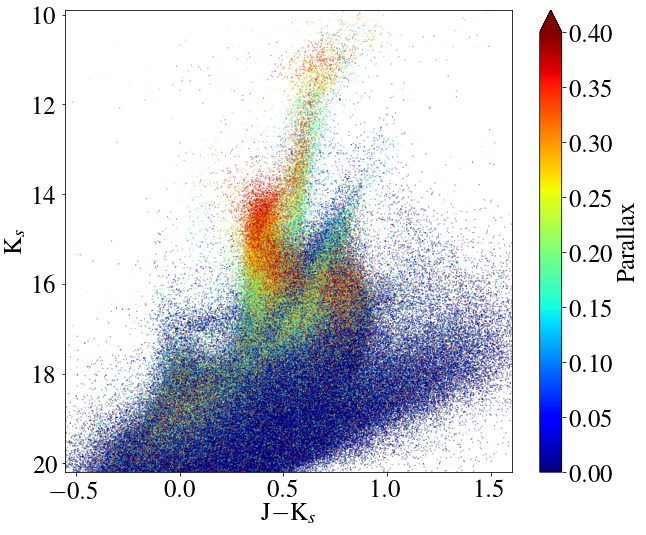}
   \includegraphics[width=\hsize]{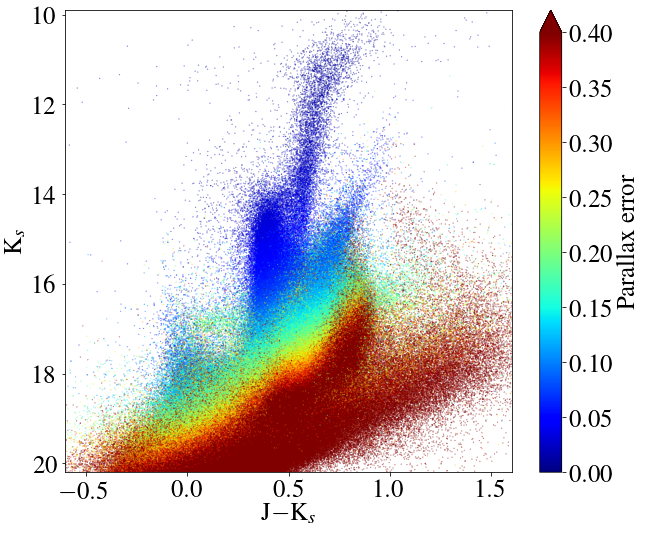}
      \caption{Near-infrared CMD of stars classified as MW foreground sources colour coded by parallax (top) and parallax error (bottom).
              }
         \label{ML_removed_MW_aux}
   \end{figure}

Figure~\ref{ML_LMC_aux} illustrates a near-infrared CMD of the initial VMC-$Gaia$~EDR3 sample compared to the training sample indicating for the latter the distribution of parallax and parallax error. This figure shows that the training sample is partly incomplete. A notable fraction of main sequence stars and carbon stars at the tip of the red giant branch (RGB) are missing from the training sample. 
However, while carbon stars are ultimately classified as LMC stars, a significant number of main sequence stars are wrongly classified as MW foreground stars along with a relatively small number of RGB stars (see Fig.~\ref{ML_removed_MW_aux} for stars classified as MW sources). These RGB stars are incorrectly classified due to the small parallax values in connection with large parallax uncertainties. The majority of properly classified bright MW stars have large parallaxes and small associated uncertainties. On the contrary, the relatively small number of incorrectly classified main sequence stars corresponds to both small parallaxes (<0.15~mas) and small parallax errors (<0.25~mas).

\subsubsection{SVM tests}
\label{tests}
%
                   

\begin{table}[] 
\caption{SVM classifier calibration}             
\label{classifier_table}      
\begin{tabular}{l c c c c}        
\hline\hline                 
& {\bf SVM} & {\bf SVM} & {\bf SVM} & {\bf SVM}\\    
& & {+ Isotonic} & {+ Sigmoid} & {$-$ parallax}\\ 
\hline                        
Brier  & 0.125 & 0.076 & 0.077 & 0.114 \\
Precision & 0.902 & 0.907 & 0.907 & 0.862\\
Recall & 0.99 & 0.985 & 0.984 & 0.984\\
\hline                                   
\end{tabular}
\end{table}

%
%



\begin{table}[]
\caption{Percentage of MW stars in different CMD regions.}
\begin{tabular}{crcccc}
\hline\hline
Region &  N & SFH$^\mathrm{a}$ &  SFH$^\mathrm{b}$ & SVM & SVM$^\mathrm{c}$ \\
\hline                     
A & 89804 & 0 & 0  &21& 40\\
B & 305702 & 0 & 0  &34& 38\\
C & 600430 & 0.3 & 1  &75& 55\\
D & 569582 & 1 & 2  &92& 67\\
E & 2335776 & 1.4 & 0.3 & 60 & 43\\
F & 316033 & 100 & 94 &95& 66\\
G & 4180 & 9.5 & 13 &12& 81\\
H & 88744 & 87 & 77 &87& 74\\
I & 180158 & 31 & 15  &52& 36\\
J & 581793 & 1.6 & 1  &32& 31\\
K & 558100 & 5.6 & 3  &23& 15\\
M & 12166 & 1.2 & 3  &4& 1\\
\hline                                   
\end{tabular}\\
$^\mathrm{(a)}$ For all tiles across the LMC (\citealp{elyoussoufi2019})\\
$^\mathrm{(b)}$ For tile LMC $8\_8$ and $13<K_\mathrm{s}<19$ mag (\citealp{Cioni2014})\\
$^\mathrm{(c)}$ Without using parallax as a parameter
\label{MWfractionTable}
\end{table}

A comparison of the previously mentioned classifier calibrations is shown in Table~\ref{classifier_table}. All three classifier calibrations provide a good predictive power, however both the Isotonic and Sigmoid calibrations are significantly closer to the best possible score of 0. Although the precision score of the Isotonic calibration is slightly lower compared to the Sigmoid calibration, the greater flexibility of the Isotonic calibration provides a better recall score. We tested several variations of the information provided to the classification algorithm and multiple tests suggest that classifications without parallax measurements are mostly unreliable. Table \ref{MWfractionTable} compares the percentage of MW stars found in the different regions of the near-infrared CMD by the SVM algorithm with and without the use of parallaxes. The latter shows that significantly fewer stars are classified as MW stars in regions H and F, which are regions clearly dominated by MW stars from star formation history (SFH) studies (e.g.~\citealp{Cioni2014,elyoussoufi2019}). In CMD regions populated mostly by faint stellar populations (A--E) the SVM classifies in general more stars as MW stars because at these magnitudes the parallax is not reliable, the proper motions are more uncertain and the efficiency of the other parameters is also reduced due to the limitation of the training sample (cf.~Fig.~\ref{MLhistograms}). There are also faint MW stars in regions I and K which are similarly difficult to isolate (see also \citealp{niederhofer2021}). The resulting percentage of LMC stars in these regions is perhaps more reliable than complete. However, the use of the parallax parameter decreases the number of MW stars in region A which contains also bright stars. This is in line with the excellent agreement between the SFH and SVM classifications of stars occupying region G.

We found that the photometric VMC data ($J$, $K_{s}$, and $J-K_{s}$) add more predictive power to the SVM classifier than the $Gaia$ photometric data ($G_{RP}$, $G_{BP}$, $G_{RP-BP}$). This is expected because the difference between the colours of MW and LMC stars is greater in the near-infrared than in the optical bands. \cite{Mazzi2021} show that the MW foreground defines a prominent feature at $J-K_{s}=0.8$ mag and a less marked one at $J-K_{s}=0.4$ mag with most of the LMC stars at $J-K_{s}<0.7$ mag. In the $Gaia$ CMD, the majority of the MW stars have colours which overlap with those of red giants in the LMC (e.g. \citealp{Vasiliev2018}). On the contrary, the VMC proper motions add no value to the predictive power compared to \textit{Gaia}~EDR3 proper motions. This is also expected because of the large uncertainties associated with the VMC proper motions of individual sources, which create less peaked distributions (Fig. \ref{MLhistograms}). 

When used on the entire {\sc StarHorse} sample (not limited to the LMC outer regions), the SVM classifier completely recovers all sources with {\sc StarHorse} distances $>$20~kpc, which includes the cleanest large sample of stars in the Magellanic Bridge \citep{schmidt2020}. Trained on the newest {\sc StarHorse} data, it also fully recovers the main sequence stars ($J-K_{s}\sim-0.2$~mag), contrary to the classifier trained on the previous {\sc StarHorse} data. Compared to the selection criteria presented in the Gaia EDR3 \citep{luri2020}, the SVM classification results in a slightly larger number of young stars ($\sim$10$^{3}$), the same number of asymptotic giant branch (AGB) stars and a significantly larger number of blue loop stars ($\sim$10$^{4}$) associated with the LMC, whereas it classifies more RGB and red clump stars ($\sim$10$^{5}$) as foreground.

\section{Results}
    \begin{table*}[]
    \caption{\textit{Gaia}~EDR3 proper motion of LMC stars. Columns (from left to right) refer to the bin number as created by the Voronoi routine, Right Ascension and Declination of the bin centre, angular distance of the bin from the LMC centre, number of stars within each bin, their average proper motion in Right Ascension and Declination, and rotational velocity with respect to the LMC centre. This table is published in its entirety (for all 293 bins) in the electronic version of the journal. The first 10 lines are shown here for guidance regarding its form and content.}
    \label{pm_table_GaiaDR2}
    \begin{tabular}{ccccccccl}
    \hline\hline
    Bin &  $\alpha$ & $\delta$ & d & N& $\mu_{\alpha}\cos{\delta}$  & $\mu_{\delta}$ & v$_\phi$\\
        &(deg)  &(deg)&(deg) &                    & (mas~yr$^{-1}$) & (mas~yr$^{-1}$)&(km~s$^{-1}$)\\
     
    \hline
1 & 84.387882 & -74.999304 & 5.19&6084 & 2.00$\pm$0.01 & 0.56$\pm$0.02 & 36.78$\pm$8.62\\
2 & 82.192321 & -74.600567 & 4.74&6699 & 2.05$\pm$0.01 & 0.45$\pm$0.02 & 40.46$\pm$8.14\\
3 & 86.752693 & -74.806033 & 5.16&5224 & 1.97$\pm$0.01 & 0.68$\pm$0.03 & 30.00$\pm$9.27\\
4 & 84.561555 & -74.419568 & 4.64&6099 & 2.01$\pm$0.01 & 0.58$\pm$0.02 & 36.10$\pm$8.64\\
5 & 87.154058 & -74.212136 & 4.66&6533 & 1.95$\pm$0.01 & 0.72$\pm$0.02 & 20.78$\pm$8.39\\
6 & 90.228859 & -74.277969 & 5.11&5067 & 1.90$\pm$0.02 & 0.83$\pm$0.03 & 20.55$\pm$9.13\\
7 & 85.705745 & -73.724596 & 4.06&5330 & 1.98$\pm$0.01 & 0.65$\pm$0.03 & 24.74$\pm$8.27\\
8 & 80.645131 & -74.078617 & 4.22&4904 & 2.05$\pm$0.02 & 0.36$\pm$0.03 & 43.73$\pm$9.98\\
9 & 88.061728 & -73.608670 & 4.24&5262 & 1.93$\pm$0.02 & 0.77$\pm$0.03 & 21.36$\pm$8.82\\
10 & 82.770292 & -73.941038 & 4.09&6460 & 2.01$\pm$0.01 & 0.48$\pm$0.02 & 32.72$\pm$8.80\\
    
    \hline
    \end{tabular}\\
    \end{table*}

    \begin{table*}[]
    \caption{\textit{Gaia} EDR3 proper motion of old stars. Columns and entire table availability are as in Table \ref{pm_table_GaiaDR2}.}
    \label{pm_table_GaiaDR2_old}
    \begin{tabular}{cccccccc}
    \hline\hline
    Bin &  $\alpha$ & $\delta$ & d &  N& $\mu_{\alpha}\cos{\delta}$     & $\mu_{\delta}$ & v$_\phi$\\
        &(deg)  &(deg)&(deg) &                    & (mas~yr$^{-1}$) & (mas~yr$^{-1}$)&(km~s$^{-1}$)\\
     
    \hline
1 & 84.370707 & -74.999242 & 5.19&5559 & 2.00$\pm$0.01 & 0.56$\pm$0.01 & 36.78$\pm$7.13\\
2 & 82.210948 & -74.602236 & 4.74&6155 & 2.05$\pm$0.01 & 0.46$\pm$0.01 & 40.46$\pm$7.16\\
3 & 86.750340 & -74.801997 & 5.16&4757 & 1.97$\pm$0.01 & 0.68$\pm$0.01 & 30.00$\pm$7.15\\
4 & 84.540673 & -74.430139 & 4.64&5552 & 2.01$\pm$0.01 & 0.58$\pm$0.02 & 36.10$\pm$6.65\\
5 & 87.148950 & -74.212158 & 4.66&6037 & 1.94$\pm$0.01 & 0.73$\pm$0.01 & 20.78$\pm$6.72\\
6 & 90.174624 & -74.267950 & 5.11&4363 & 1.90$\pm$0.01 & 0.83$\pm$0.01 & 20.55$\pm$7.74\\
7 & 85.712817 & -73.724449 & 4.06&4842 & 1.98$\pm$0.01 & 0.65$\pm$0.01 & 24.74$\pm$7.31\\
8 & 80.646723 & -74.075803 & 4.22&4543 & 2.05$\pm$0.01 & 0.36$\pm$0.02 & 43.73$\pm$7.95\\
9 & 88.069836 & -73.610031 & 4.24&4793 & 1.93$\pm$0.01 & 0.78$\pm$0.02 & 21.36$\pm$8.09\\
10 & 82.753622 & -73.946524 & 4.09&5862 & 2.01$\pm$0.01 & 0.48$\pm$0.02 & 32.72$\pm$8.08\\
    
    \hline
    \end{tabular}\\
    \end{table*}
    
    \begin{table*}[]
    \caption{\textit{Gaia} EDR3 proper motion of young stars. Columns and entire table availability are as in Table \ref{pm_table_GaiaDR2}.}
    \label{pm_table_GaiaDR2_young}
    \begin{tabular}{ccccrrrr}
    \hline\hline
    Bin &  $\alpha$ & $\delta$ & d & N & $\mu_{\alpha}\cos{\delta}$     & $\mu_{\delta}$ & v$_\phi$\\
        & (deg) &(deg)&(deg) &                & (mas~yr$^{-1}$) & (mas~yr$^{-1}$)&(km~s$^{-1}$)\\
     
    \hline
1 & 84.412245 & -75.015064 & 5.21&363 & 2.01$\pm$0.08 & 0.54$\pm$0.14 & 41.59$\pm$25.52\\
2 & 82.322068 & -74.590725 & 4.73&424 & 2.05$\pm$0.06 & 0.43$\pm$0.08 & 45.95$\pm$19.59\\
3 & 86.811303 & -74.833736 & 5.20&339 & 2.00$\pm$0.12 & 0.69$\pm$0.14 & 35.76$\pm$34.65\\
4 & 84.501116 & -74.377691 & 4.59&369 & 2.01$\pm$0.08 & 0.60$\pm$0.14 & 32.86$\pm$23.68\\
5 & 87.289344 & -74.209547 & 4.67&297 & 2.00$\pm$0.08 & 0.73$\pm$0.23 & 33.58$\pm$23.67\\
6 & 90.474877 & -74.391194 & 5.26&422 & 1.91$\pm$0.06 & 0.84$\pm$0.30 & 21.83$\pm$19.26\\
7 & 85.894985 & -73.744076 & 4.09&331 & 2.01$\pm$0.10 & 0.66$\pm$0.35 & 31.26$\pm$27.61\\
8 & 80.690214 & -74.092922 & 4.23&230 & 2.04$\pm$0.06 & 0.35$\pm$0.11 & 43.58$\pm$20.60\\
9 & 87.950156 & -73.560522 & 4.18&315 & 1.96$\pm$0.05 & 0.79$\pm$0.13 & 29.23$\pm$15.74\\
10 & 82.783582 & -73.986885 & 4.13&338 & 2.03$\pm$0.11 & 0.50$\pm$0.10 & 33.47$\pm$32.54\\
    
    \hline
    \end{tabular}\\
    \end{table*}
    
    \begin{table*}[]
    \caption{VMC proper motion of LMC stars. Columns and entire table availability are as in Table \ref{pm_table_GaiaDR2}.}
    \label{pm_table_VMC}
    \begin{tabular}{cccccrrr}
    \hline\hline
    Bin &  $\alpha$ & $\delta$ & d & N& $\mu_{\alpha}\cos{\delta}$     & $\mu_{\delta}$ & v$_\phi$\\
        & (deg) &(deg)&(deg) &                    & (mas~yr$^{-1}$) & (mas~yr$^{-1}$)&(km~s$^{-1}$)\\
     
    \hline
1 & 84.387882 & -74.999304 & 5.19&6090 & 2.01$\pm$0.03 & 0.57$\pm$0.04 & 37.19$\pm$12.16\\
2 & 82.192321 & -74.600567 & 4.73&7947 & 2.06$\pm$0.03 & 0.39$\pm$0.03 & 41.35$\pm$11.34\\
3 & 86.752693 & -74.806033 & 5.17&6153 & 1.94$\pm$0.03 & 0.67$\pm$0.04 & 30.21$\pm$12.62\\
4 & 84.561555 & -74.419568 & 4.63&6335 & 2.01$\pm$0.03 & 0.60$\pm$0.04 & 35.87$\pm$12.12\\
5 & 87.154058 & -74.212136 & 4.66&7725 & 1.94$\pm$0.02 & 0.70$\pm$0.03 & 21.05$\pm$11.69\\
6 & 90.228859 & -74.277969 & 5.13&5690 & 1.89$\pm$0.03 & 0.81$\pm$0.04 & 19.43$\pm$13.28\\
7 & 85.705745 & -73.724596 & 4.06&6043 & 1.97$\pm$0.03 & 0.65$\pm$0.04 & 24.81$\pm$12.78\\
8 & 80.645131 & -74.078617 & 4.22&4909 & 2.06$\pm$0.04 & 0.34$\pm$0.04 & 44.47$\pm$13.75\\
9 & 88.061728 & -73.608670 & 4.24&5265 & 1.93$\pm$0.04 & 0.74$\pm$0.04 & 21.51$\pm$14.24\\
10 & 82.770292 & -73.941038 & 4.09&7642 & 1.95$\pm$0.03 & 0.46$\pm$0.03 & 32.44$\pm$11.71\\

    \hline
    \end{tabular}\\
    \end{table*}

To study the kinematics of the outer regions of the LMC, we used a sample of 2,629,456 LMC stars, obtained after removing MW stars as described above. Figure \ref{cmd_populations} shows the distribution of these stars in the near-infrared CMD and regions occupied by specific stellar populations are indicated as in \cite{elyoussoufi2019}. We used the two-dimensional Voronoi technique  \citep{Cappellari2003}, as in \cite{schmidt2020}, to bin the data in sky coordinates. We used 257 bins where each bin contains from a hundred to several thousands LMC stars, depending on the selected stellar population as described in Section \ref{spop}. 
We derived the median \textit{Gaia} EDR3 and VMC proper motions and the corresponding uncertainties of each bin. These values are listed in Tables~\ref{pm_table_GaiaDR2}$-$\ref{pm_table_VMC}. which are available in its entirety only electronically. The first ten lines show that for each bin we provide its angular distance from the centre of the LMC \citep{paturel2003}, the number of stars, the proper motion components and the rotational velocity derived from both the \textit{Gaia} EDR3 and VMC data. Although we can measure a significantly larger number of proper motions from the VMC survey alone, this advantage is lost due to the requirement of an existing parallax measurement for the SVM algorithm, which we used to remove the MW stars.

   \begin{figure}
   \centering
   \includegraphics[width=\hsize]{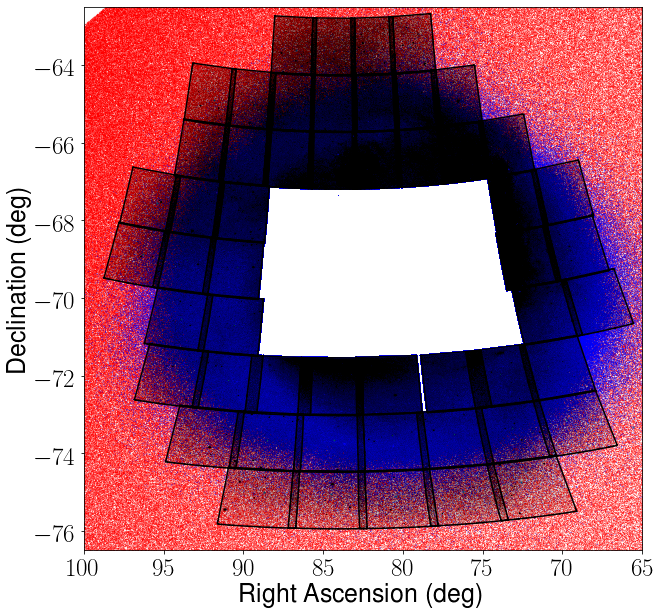}
      \caption{VMC tiles (black) covering the outer regions of the LMC superimposed on the distribution of \textit{Gaia} EDR3 mock data. MW foreground stars in the mock catalogue are shown in red and LMC stars in blue.
              }
         \label{mock_overlap}
   \end{figure}

We used the LMC population of a \textit{Gaia} EDR3 mock catalogue of \cite{rybizki2020} as a tool to derive the expected proper motion measurements across the outer LMC for a non-rotating spherical galaxy with no internal kinematics. 
The mock catalogue is constructed as a simple 3-dimensional (3D) Gaussian distribution. All the particles are assigned a random 3D velocity centred around the bulk motion of the LMC which corresponds to 1.95~mas yr$^{-1}$ in Right Ascension, 0.43~mas yr$^{-1}$ in Declination, and $v_{R}=$283~km s$^{-1}$ for the radial velocity corrected for the solar reflex motion with respect to the Galactic centre. The mock catalogue proper motions are then derived from the position on the sky, the distance and 3D velocity. The mock catalogue contains among other things a Magellanic Clouds stellar population, which can be queried by its identification number ($popid$ = 10). The sky position of the centre-of-mass of the LMC model is described in \cite{paturel2003} and corresponds to 80$^\circ$.84 in Right Ascension and $-$69$^\circ$.78 in Declination. The overlap between the mock catalogue and the VMC tiles is shown in Fig.~\ref{mock_overlap}; the LMC footprint from the VMC survey is entirely covered. As mentioned by \citet{rybizki2020}, the proper motions in the model data should not be compared to the \textit{Gaia} measurements, since the model does not account for the internal kinematics and components of the galaxy (e.g. disc, rotation, spiral arms and bar). However, the model data can be used to infer a rotation field which proves useful in our study. Furthermore, it allows us to correct for the changes in proper motion due to the 3D bulk motion of the LMC (\citealp{robin2012}) caused by different viewing angles. We took the shape and size of our Voronoi bins into account and transferred those onto the model. The different viewing angles are the only cause of any difference in the median proper motion of these model bins from the central proper motion. We then subtracted the median predicted proper motion of the model bins from the measured median proper motions to remove the LMC bulk motion. This enables studying the internal kinematics of the LMC (e.g. rotation), this also corrects the measurements for the differences in viewing angles and the solar reflex motion. Without applying such a correction, the proper motions, especially to the east and west of the LMC, appear significantly higher than overall around the galaxy.

Furthermore, the proper motion values in the mock catalogue are already corrected for the reflex motion of the Sun\footnote{https://dc.zah.uni-heidelberg.de/browse/gedr3mock/q}. The space velocity of the Sun adopted for this correction corresponds to the Galactocentric velocity ($v_{x}$=11.1~km~s$^{-1}$, $v_{y}$=239.08~km~s$^{-1}$, $v_{z}$=7.25~km~s$^{-1}$), which is different from the latest value derived using \textit{Gaia} EDR3 data with respect to compact (quasar-like) extragalactic sources \citep{klioner2020}. The solar reflex motion influences proper motion measurements depending on the distance of the sources and their position on the sky. While the effect diminishes with distance, it is notable at the distance of the LMC (Fig.~\ref{reflex_motion_vsun_model}). Compared to the \textit{Gaia} EDR3 values there is an average offset in the proper motion of 0.01~mas~yr$^{-1}$ in Declination and of 0.02~mas~yr$^{-1}$ in Right Ascension (Fig.~\ref{reflex_motion}). These offsets are at least ten times smaller than the average measurement uncertainties.

\subsection{Selection of stellar populations}\label{spop}
   \begin{figure}
   \centering
   \includegraphics[width=\hsize]{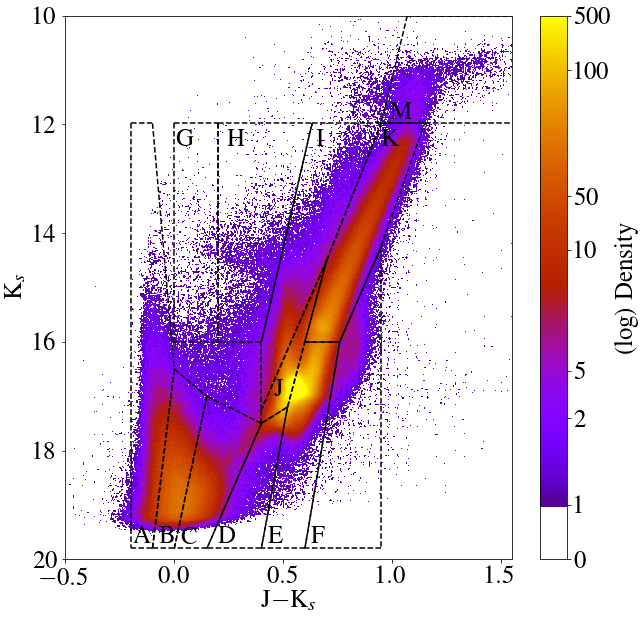}
      \caption{Near-infrared CMD of LMC stars with boxes overlaid to distinguish different populations according to \cite{elyoussoufi2019}.
              }
         \label{cmd_populations}
   \end{figure}

We used the CMD regions from \citet{elyoussoufi2019} to select different stellar populations (see Fig. \ref{cmd_populations}). Each region is dominated by a class of objects with an associated median age (see table 1 in \citealp{elyoussoufi2019}). 
Compared to \citet{elyoussoufi2019}, we find significantly fewer MW foreground stars, especially in region F. Except for stars at the edge of the RGB, this region contains mainly MW foreground sources wrongly classified as LMC sources by the machine learning algorithm. Sources in region F are excluded from the subsequent analysis. Using the remaining CMD regions, we divided the catalogue into two age groups: a young population ($<$1 Gyr old) containing stars in the CMD regions A, B, G, H and I and an older population ($>$2 Gyr old) which contains stars in the CMD regions D, E, J, K and M. This age division is based on average ages of the CMD bins as described in \citet{elyoussoufi2019}. We do not use stars in region C because they have ages encompassing both groups. There are 808,515 and 2,738,707 stars in the young and old age group, respectively.

\subsection{Proper motion maps} \label{LMC_kinematics}
   \begin{figure}
   \centering
   \includegraphics[width=\hsize]{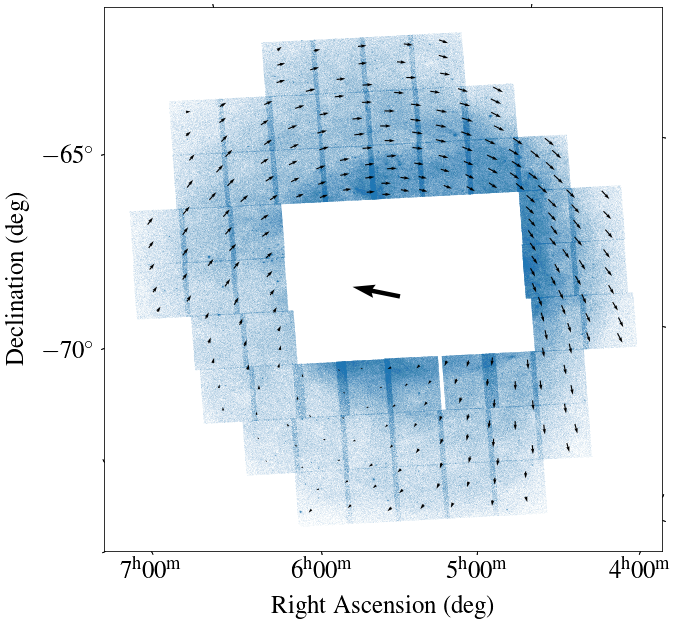}
      \caption{Distribution of stellar proper motions across the LMC from the \textit{Gaia} EDR3 sample. Each vector represents the median proper motion of at least a few hundred stars. The central vector indicates the bulk motion at the centre of the GalMod model, that was used to create the residual map. The central regions which are characterised by a high stellar density and influenced by crowding issues are not analysed in this study.
              }
         \label{LMC_rotaion_GaiaDR2}
   \end{figure}

   \begin{figure}
   \centering
   \includegraphics[width=\hsize]{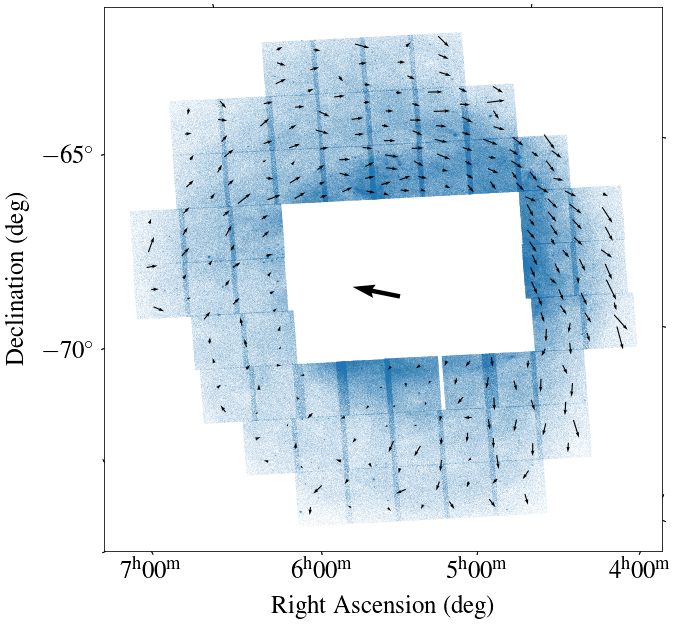}
      \caption{As Fig.~\ref{LMC_rotaion_GaiaDR2} but for VMC proper motions. Each vector represents the median proper motion of at least 1600 stars.}
         \label{LMC_rotaion_VMC}
   \end{figure}
%
%

\begin{figure}
   \centering
   \includegraphics[width=\hsize]{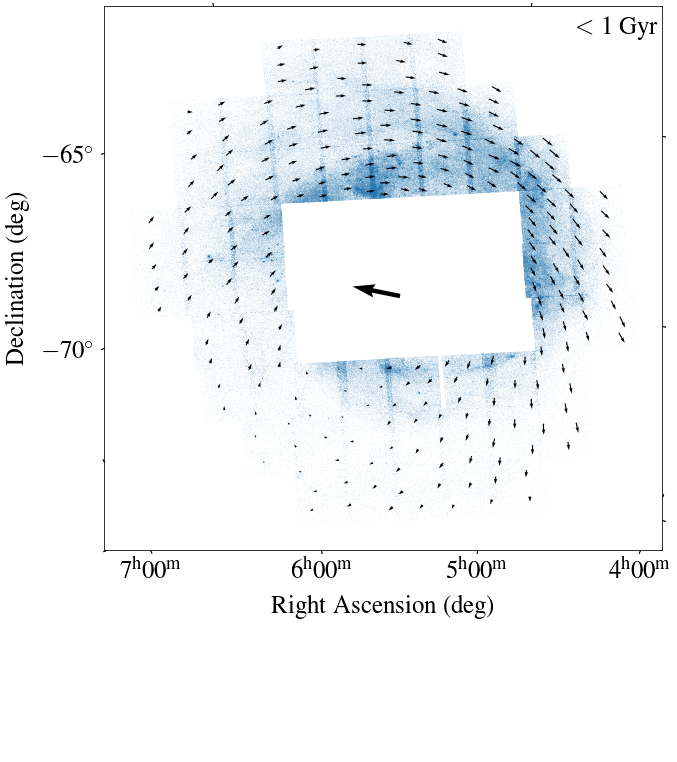}
   \includegraphics[width=\hsize]{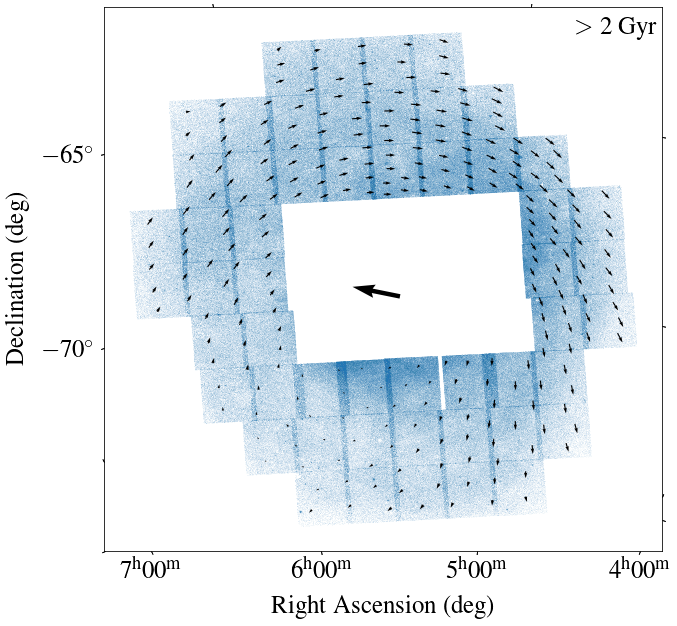}
      \caption{\textit{Gaia} EDR3 proper motion maps of the LMC for samples of stars $<$1~Gyr old (top) and $>$2~Gyr old  (bottom), see text for details. Each vector represents the median proper motion of at least 100 stars for the young population and of at least 2000 stars for the old population. The central vector indicates the bulk motion of the LMC as shown in Fig.\ref{LMC_rotaion_GaiaDR2}.
      }
         \label{populations}
   \end{figure}

\begin{figure}
   \centering
   \includegraphics[width=\hsize]{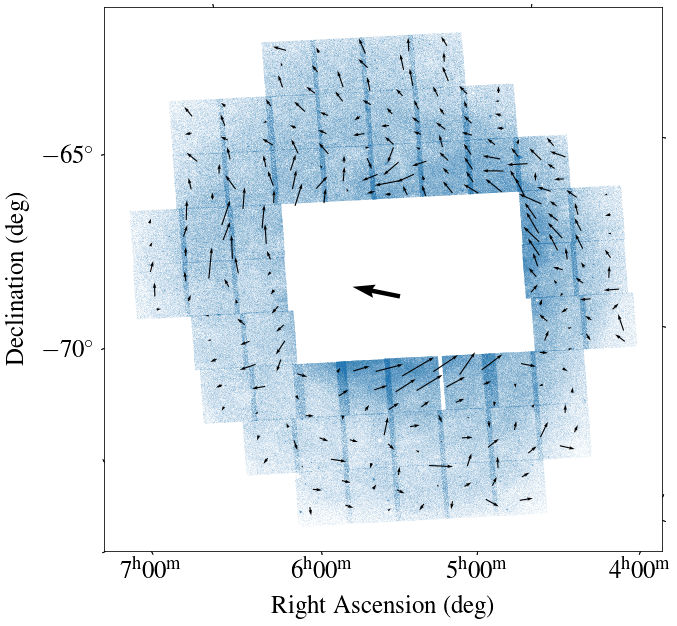}
      \caption{\textit{Gaia} EDR3 differential proper motion map of the LMC. With the proper motion of the young population (Fig.~\ref{populations} top) subtracted from that of the old population (Fig.~\ref{populations} bottom). The proper motion vectors are up scaled by a factor of ten compared to Fig.\ref{populations}. 
      }
         \label{old_minus_young}
   \end{figure}
%
%

  Figures \ref{LMC_rotaion_GaiaDR2} and \ref{LMC_rotaion_VMC} show the residual proper motion maps of LMC stars from the \textit{Gaia} EDR3 and VMC samples respectively, after subtracting the LMC bulk motion. Both maps show rotation of the LMC as expected. The rotation is clearly visible in the northern and western regions of the outer LMC, where the northwestern and western spiral arms are located. Regions in the east and parts of the south-west also show a visible rotation. Only the south eastern region, where the south eastern arm is located, does not show a clear rotation compared to the rest of the galaxy. The residual proper motions are significantly smaller. However, the proper motions of most bins in this regions follow the expected direction of the LMC rotation. 
  
  Figure \ref{populations} shows residual proper motion maps of our selected stellar populations: $<$1~Gyr and $>$2~Gyr old (see Section \ref{spop}). The old population, which consists of about two thirds of all LMC stars, shows a smooth rotation especially in the north. The young population shows a similar behaviour, but for a less coherent motion in the region of the northern arm, where the young population is likely influenced by the morphological structures of the LMC. To better compare the proper motion of old and young stars we show their differential map (Fig.~\ref{old_minus_young}). The differential motion shows a few distinct aspects. Residuals on the northern and north-eastern sides point predominantly towards the North, which could be due to the influence of the MW after the pericentre passage (\citealp{vanderMarel2002}) as this is the direction towards the MW centre. Residuals on the south and south-eastern side do not show any preferential motion which suggests that young and old stars share the same motion. However, residuals in the inner region south, west and north of the bar show a counter-clockwise rotation which indicates that young stars move faster than old stars.

 \subsection{Rotation}
   \begin{figure}
   \centering
   \includegraphics[width=\hsize]{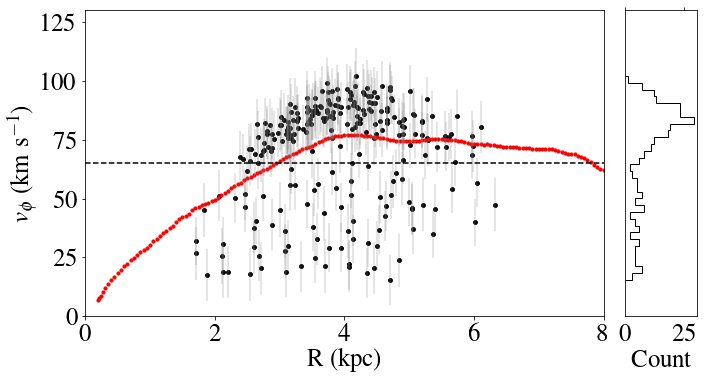}
      \caption{Rotational velocity of the LMC derived from \textit{Gaia} EDR3 proper motions with respect to the centre of the \textit{Gaia} EDR3 mock catalogue (black). Each point corresponds to the median velocity within bins in the sky that contain at least a few hundred stars (see text for detail).The red line indicates the \textit{Gaia} EDR3 measurements \citep{luri2020}. The dashed line at 65~km~s$^{-1}$ was chosen based on the histogram to the right of the main diagram.}
         \label{rotation_curve_GaiaDR2}
   \end{figure}

We used the python package $galpy$ to derive the rotational velocity component with respect to the LMC centre in km~s$^{-1}$ (v$_\phi$) and the median distance to the centre of each Voronoi bin. For the centre we adopted the values 80$^\circ$.84 in Right Ascension and $-$69$^\circ$.78 in Declination, which is the value used to produce the \textit{Gaia} EDR3 mock catalogue \citep{rybizki2020}.
The rotational velocities calculated from all stars within each bin are shown in Fig.~\ref{rotation_curve_GaiaDR2}. The majority of the data points show velocities higher than the \textit{Gaia} EDR3  (e.g.~\citealp{luri2020}) rotation curve, but a non-negligible fraction shows below-average rotation speeds. Nearly all bins corresponding to high speeds are located in the north-east to Western regions of the LMC. The difference between the higher speeds in this study and the \textit{Gaia}~EDR3 measurements are explained by the different binning techniques. The larger sample size in this study eliminates the need for radial binning and therefore enables a more differentiated map of the outer LMC. The velocities for young and old stars show no significant differences from the combined sample. The younger population shows however a larger spread, due to the larger uncertainties. Similar to their combined sample the majority of both the young and old stars show a slightly higher rotation speed compared to the \textit{Gaia} EDR3 results. 

Contrary to radial binning, which is often used to measure rotation curves, our binning enables a more differentiated map of the outer LMC. 

   \begin{figure}
   \centering
   \includegraphics[width=\hsize]{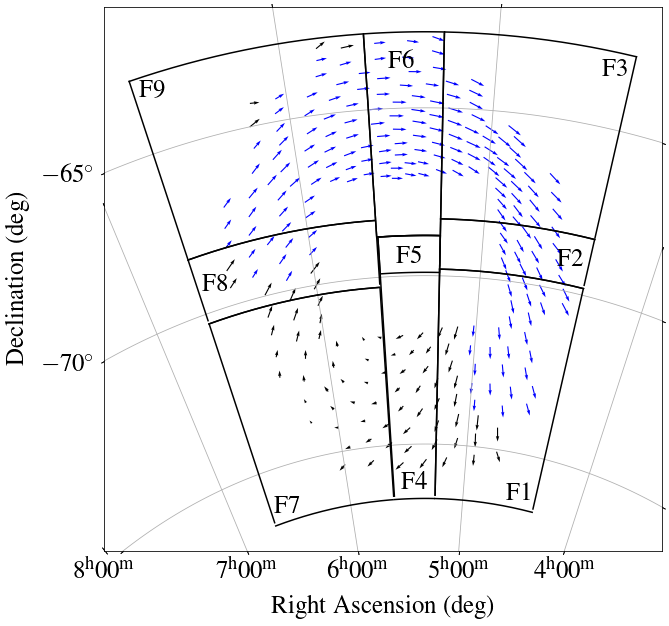}
        \caption{{\textit{Gaia}~EDR3} proper motion map of the LMC stars in bins (containing each at least a few thousand stars from different stellar populations) colour coded by rotational velocity, black proper motion vectors indicate a rotation speed of less than 65~km~s$^{-1}$ (dashed line in Fig.~\ref{rotation_curve_GaiaDR2}). Regions F1$-$9 are from \cite{parada2021}.
        }
        \label{LMC_rotaion_GaiaDR2_colour_coded}%
   \end{figure}

Figure~\ref{LMC_rotaion_GaiaDR2_colour_coded} shows the median proper motions derived for all stars within the Voronoi bins colour-coded by their rotational speed with superimposed the spatial regions defined in \cite{parada2021}. 
These authors illustrated the 3D orientation of the LMC by introducing nine (F1$-$9) spatial regions. They used carbon stars as standard candles to infer distances from their median $J$ magnitudes. For the LMC they found the region F1 ($\Delta \overline{M}_{J}=$~0.11~mag) to be the furthest away and F9 ($\Delta \overline{M}_{J}=-$0.12~mag) to be the closest, while F7 ($\Delta \overline{M}_{J}=$~0.00~mag), F2/F4($\Delta \overline{M}_{J}=$~0.01~mag), F3/F6 ($\Delta \overline{M}_{J}=-$0.04~mag) and F5 ($\Delta \overline{M}_{J}=-$0.02~mag) are at a distance similar to the average distance. Only region F8 ($\Delta \overline{M}_{J}=$~0.07~mag) deviated from expectations, but this was explained by the high extinction surrounding the 30 Doradus star forming region.

The resulting pattern suggests that a part of the LMC (south east) is rotating slower than expected, which is apparently not correlated with the orientation of the galaxy. We recall that our proper motion maps are not corrected for the inclination of the LMC disc, which may introduce a velocity pattern.
\begin{figure*}
   \centering
   \includegraphics[width=1.0\textwidth]{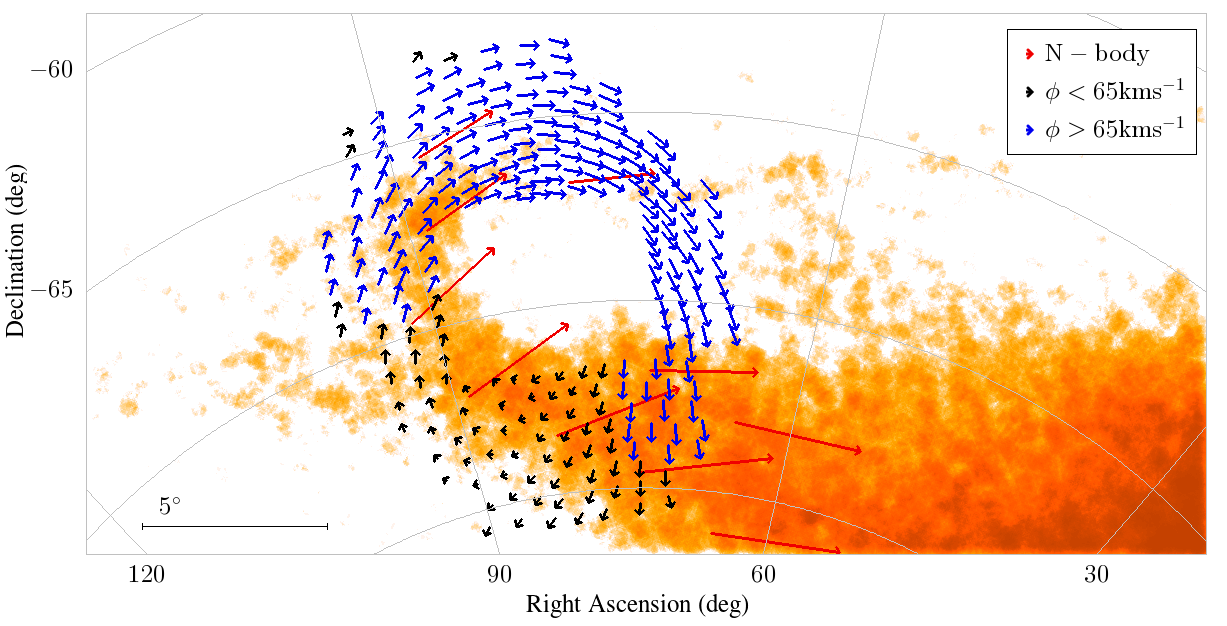}
   
   \caption{\textit{Gaia}~EDR3 proper motions of the full sample of LMC stars, without distinguishing between young and old stellar populations, superimposed on the model density distribution of stripped SMC stars \citep{diaz2012}. The proper motion vectors are colour coded by the rotational velocity with stars rotating slower than 65 km s$^{-1}$ indicated in black and those rotating faster indicated in blue. Each arrow corresponds to the ensemble velocity of at least a few thousand stars. The red arrows indicate the median proper motion of the $N$-body particles.}
    \label{LMC_rotaion_GaiaDR2_SMC_nbody}%
    \end{figure*}
One possible explanation would involve a population of stars connected to the debris of the SMC in the south-eastern regions (F4, F7 and F8) of the LMC, which resulted from the last interaction event between the two galaxies (roughly 250 Myr ago). The origin of the material is unclear. It can be stripped SMC material, LMC material influenced by the SMC or more likely a combination of both. An $N$-body simulation finds this event responsible for pulling the Bridge out of the SMC disc \citep{diaz2012}.
This debris population apparently connects towards both the young (F1) and old (F4) Bridge and possibly bends around the LMC on the northeast (F9). The connection towards the Magellanic (young) Bridge and the Southern Substructure (old Bridge) coincides with the locations found by \citet{Belokurov2017} and \citet{ElYoussoufi2021}.


\subsection{Influence of the SMC impact}
To interpret the velocity pattern of the LMC stars, we compare the map shown in Fig.~\ref{LMC_rotaion_GaiaDR2} with an $N$-body model of the LMC--SMC system from \cite{diaz2012}. For this model the authors adopted an orbital history in which the LMC and the SMC have become an interacting pair since about 2 Gyr. As a result of their first encounter, the SMC is disrupted, several tidal structures are created and a large fraction of the material stripped away from the SMC is engulfed by the LMC. 
Figure \ref{LMC_rotaion_GaiaDR2_SMC_nbody} shows the 
proper motion of the LMC stars  superimposed on the density distribution of the $N$-body simulation. The motion of the $N$-body particles (indicated with red arrows) is mainly towards the West and the predicted distribution of the SMC debris spans the area from the South-West to the North-East of the galaxy. The slower rotational speed of LMC stars in the South-East compared to those in other areas, could result from the influence of the SMC.

A kinematically distinct component due to the SMC was found by \cite{olsen2011, olsen2015} from the analysis of about 6000 spectra of giant stars. The authors derived both radial velocities and metallicities to conclude that about 5\% of their stellar sample originated in the SMC. These stars after being subsequently captured by the LMC, together with gas, might have triggered the intense star formation in 30 Doradus region. The rotational pattern of these suspected SMC stars opposes the LMC rotation by more than 50~km~s$^{-1}$. This velocity translates to a proper motion of about 0.2 mas~yr$^{-1}$ at the distance of the LMC, which is within the uncertainties of both our VMC and \textit{Gaia}~EDR3 proper motion measurements.
However, assuming we have detected the influence of the SMC stars in the South-East of the LMC, this suggests that either the velocity difference is larger than 50~km~s$^{-1}$ or that the fraction of stars sharing this velocity is significantly larger. Assuming that these stars move as predicted by the $N$-body simulation (\citealp{diaz2012}) we would expect a fraction of about 6\%, which is in good agreement with the results found by \citet{olsen2015}. It is interesting that the percentages of kinematically distinct stars agree despite the differences between the two stellar samples. Our sample includes a broad range of stellar populations: from main sequence to red clump stars, the bright part of the RGB and other evolved giants (Fig.~\ref{cmd_populations}). The sample used by \citet{olsen2015} is limited to $K_{s}<12$ mag and contains predominantly massive red supergiants and AGB stars, including carbon stars. 
 
\cite{indu2015} modelled the neutral atomic hydrogen kinematics of the LMC and found that about 12\% of the data points deviate from that of a disc. These points trace various features, possibly caused by tidal and/or hydrodynamical effects. In the South-East of the galaxy the gas is rotating slower than in the main disc which is similar to what we find in our study. \cite{indu2015} suggest that an infall of gas from the Magellanic Bridge, at the South-West, would explain the counter rotating component identified by \cite{olsen2011} and the slow rotation in the South-East, the latter resulting from the drag of the rotating component.

The tidal force of the MW during the LMC infall could also influence the rotational speed by stretching the LMC in a direction perpendicular to the bar. However, this effect does not seem to be very strong, especially in the southern regions of the galaxy (Fig.~\ref{old_minus_young}). A better understanding of the particular region of the LMC influenced by the recent SMC pericentre ($\sim$ 150 Myr ago) and the disc crossings (400 Myr ago and older), see for example \cite{cullinane2022} for the location of these events, as well as a separation of the LMC and SMC stars will be crucial to explain the kinematic properties of the galaxy.


\section{Conclusions}
\label{ch:Conclusions}

In this study we continued our investigation of the outer regions of the Magellanic system. We combined the VMC data with data from the $Gaia$~EDR3 to study the outer regions of the LMC. We introduced a new method to distinguish between Magellanic and MW stars, based on a machine learning algorithm trained on {\sc StarHorse} distance estimates, to improve our analysis by significantly increasing the usable sample size. Our membership classification provides the largest catalogue of LMC stars with a distance between 1.5 and 6.2 kpc from the centre of the galaxy. With this technique we are able to study the kinematics of faint stellar populations below the RC of the LMC. We showed residual proper motion maps (after subtracting the bulk motion) of two populations (<1 and >2 Gyr old). Both populations showed a  distinct motion in the South--East outer region of the galaxy compared to the other directions. For the first time we found hints of stripped SMC debris in the proper motion of stars located in south east of the outer LMC. This supports findings by \citet{olsen2011} that suggested a counter-rotating population in that region using radial velocities of giant stars. The current uncertainties of the proper motions are however insufficient to separate those stars from possible LMC stars for which their motion might have been influenced by  interaction with the SMC.

In the future, projects like the 4MOST will provide a large sample of radial velocities for different stellar populations across the Magellanic Clouds \citep{4mostCioni2019}, that are currently missing, to determine the fraction of SMC stars accreted by the LMC. It may also be possible to determine if individual stars belong to the LMC or the SMC by chemical tagging. So far, it is unclear if these stars, that rotate more slowly than the average rotational speed in the outer region of the LMC are left-overs from the SMC interaction or if the rotational direction of LMC stars has changed as a result of the interaction between the two galaxies. 

\begin{acknowledgements} 
This project has received funding from the European Research Council (ERC) under the European Union’s Horizon 2020 research and innovation programme (grant agreement no. 682115). We thank the Cambridge Astronomy Survey Unit (CASU) and the Wide Field Astronomy Unit (WFAU) in Edinburgh for providing the necessary data products under the support of the Science and Technology Facility Council (STFC) in the UK. This research was also supported in part by the Australian Research Council Centre of Excellence for All Sky Astrophysics in 3 Dimensions (ASTRO 3D), through project number CE170100013.
This project has made extensive use of the Tool for OPerations on Catalogues And Tables (TOPCAT) software package (Taylor 2005) as well as the following open-source Python packages: Astropy (The Astropy Collaboration et al.2018), matplotlib (Hunter 2007), NumPy (Oliphant 2015), Numba, and Pyraf. \end{acknowledgements}

\bibliographystyle{aa} 
\bibliography{bibtex.bib}

\begin{thebibliography}{61}
\expandafter\ifx\csname natexlab\endcsname\relax\def\natexlab#1{#1}\fi

\bibitem[{{Anders} {et~al.}(2021){Anders}, {Khalatyan}, {Queiroz}, {Chiappini},
  {Ard{\`e}vol}, {Casamiquela}, {Figueras}, {Jim{\'e}nez-Arranz}, {Jordi},
  {Mongui{\'o}}, {Romero-G{\'o}mez}, {Altamirano}, {Antoja}, {Assaad},
  {Cantat-Gaudin}, {Castro-Ginard}, {Enke}, {Girardi}, {Guiglion}, {Khan},
  {Luri}, {Miglio}, {Minchev}, {Ramos}, {Santiago}, \&
  {Steinmetz}}]{Anders2021}
{Anders}, F., {Khalatyan}, A., {Queiroz}, A.~B.~A., {et~al.} 2021, arXiv
  e-prints, arXiv:2111.01860

\bibitem[{{Belokurov} {et~al.}(2017){Belokurov}, {Erkal}, {Deason}, {Koposov},
  {De Angeli}, {Evans}, {Fraternali}, \& {Mackey}}]{Belokurov2017}
{Belokurov}, V., {Erkal}, D., {Deason}, A.~J., {et~al.} 2017, \mnras, 466, 4711

\bibitem[{{Besla} {et~al.}(2013){Besla}, {Hernquist}, \& {Loeb}}]{besla2013}
{Besla}, G., {Hernquist}, L., \& {Loeb}, A. 2013, \mnras, 428, 2342

\bibitem[{{Besla} {et~al.}(2007){Besla}, {Kallivayalil}, {Hernquist},
  {Robertson}, {Cox}, {van der Marel}, \& {Alcock}}]{Besla2007}
{Besla}, G., {Kallivayalil}, N., {Hernquist}, L., {et~al.} 2007, \apj, 668, 949

\bibitem[{{Besla} {et~al.}(2016){Besla}, {Mart{\'\i}nez-Delgado}, {van der
  Marel}, {Beletsky}, {Seibert}, {Schlafly}, {Grebel}, \& {Neyer}}]{Besla2016}
{Besla}, G., {Mart{\'\i}nez-Delgado}, D., {van der Marel}, R.~P., {et~al.}
  2016, \apj, 825, 20

\bibitem[{Boser {et~al.}(1992)Boser, Guyon, \& Vapnik}]{Boser1992}
Boser, B.~E., Guyon, I.~M., \& Vapnik, V.~N. 1992, in Proceedings of the Fifth
  Annual Workshop on Computational Learning Theory, COLT '92 (New York, NY,
  USA: Association for Computing Machinery), 144–152

\bibitem[{{Cappellari} \& {Copin}(2003)}]{Cappellari2003}
{Cappellari}, M. \& {Copin}, Y. 2003, MNRAS, 342, 345

\bibitem[{{Cioni} {et~al.}(2019){Cioni}, {Storm}, {Bell}, {Lemasle},
  {Niederhofer}, {Bestenlehner}, {El Youssoufi}, {Feltzing},
  {Gonz{\'a}lez-Fern{\'a}ndez}, {Grebel}, {Hobbs}, {Irwin}, {Jablonka}, {Koch},
  {Schnurr}, {Schmidt}, \& {Steinmetz}}]{4mostCioni2019}
{Cioni}, M. . R.~L., {Storm}, J., {Bell}, C.~P.~M., {et~al.} 2019, The
  Messenger, 175, 54

\bibitem[{{Cioni} {et~al.}(2016){Cioni}, {Bekki}, {Girardi}, {de Grijs},
  {Irwin}, {Ivanov}, {Marconi}, {Oliveira}, {Piatti}, {Ripepi}, \& {van
  Loon}}]{cioni2016}
{Cioni}, M.-R.~L., {Bekki}, K., {Girardi}, L., {et~al.} 2016, \aap, 586, A77

\bibitem[{{Cioni} {et~al.}(2011){Cioni}, {Clementini}, {Girardi}, {Guand
  alini}, {Gullieuszik}, {Miszalski}, {Moretti}, {Ripepi}, {Rubele}, {Bagheri},
  {Bekki}, {Cross}, {de Blok}, {de Grijs}, {Emerson}, {Evans}, {Gibson},
  {Gonzales-Solares}, {Groenewegen}, {Irwin}, {Ivanov}, {Lewis}, {Marconi},
  {Marquette}, {Mastropietro}, {Moore}, {Napiwotzki}, {Naylor}, {Oliveira},
  {Read}, {Sutorius}, {van Loon}, {Wilkinson}, \& {Wood}}]{cioni2011}
{Cioni}, M. R.~L., {Clementini}, G., {Girardi}, L., {et~al.} 2011, \aap, 527,
  A116

\bibitem[{{Cioni} {et~al.}(2014){Cioni}, {Girardi}, {Moretti}, {Piffl},
  {Ripepi}, {Rubele}, {Scholz}, {Bekki}, {Clementini}, {Ivanov}, {Oliveira}, \&
  {van Loon}}]{Cioni2014}
{Cioni}, M. R.~L., {Girardi}, L., {Moretti}, M.~I., {et~al.} 2014, \aap, 562,
  A32

\bibitem[{{Cross} {et~al.}(2012){Cross}, {Collins}, {Mann}, {Read}, {Sutorius},
  {Blake}, {Holliman}, {Hambly}, {Emerson}, {Lawrence}, \&
  {Noddle}}]{Cross2012}
{Cross}, N.~J.~G., {Collins}, R.~S., {Mann}, R.~G., {et~al.} 2012, \aap, 548,
  A119

\bibitem[{{Cullinane} {et~al.}(2022){Cullinane}, {Mackey}, {Da Costa}, {Erkal},
  {Koposov}, \& {Belokurov}}]{cullinane2022}
{Cullinane}, L.~R., {Mackey}, A.~D., {Da Costa}, G.~S., {et~al.} 2022, \mnras,
  510, 445

\bibitem[{{de Jong} {et~al.}(2019){de Jong}, {Agertz}, {Berbel}, {Aird},
  {Alexander}, {Amarsi}, {Anders}, {Andrae}, {Ansarinejad}, {Ansorge},
  {Antilogus}, {Anwand-Heerwart}, {Arentsen}, {Arnadottir}, {Asplund}, {Auger},
  {Azais}, {Baade}, {Baker}, {Baker}, {Balbinot}, {Baldry}, {Banerji},
  {Barden}, {Barklem}, {Barth{\'e}l{\'e}my-Mazot}, {Battistini}, {Bauer},
  {Bell}, {Bellido-Tirado}, {Bellstedt}, {Belokurov}, {Bensby}, {Bergemann},
  {Bestenlehner}, {Bielby}, {Bilicki}, {Blake}, {Bland-Hawthorn}, {Boeche},
  {Boland}, {Boller}, {Bongard}, {Bongiorno}, {Bonifacio}, {Boudon}, {Brooks},
  {Brown}, {Brown}, {Br{\"u}ggen}, {Brynnel}, {Brzeski}, {Buchert},
  {Buschkamp}, {Caffau}, {Caillier}, {Carrick}, {Casagrande}, {Case}, {Casey},
  {Cesarini}, {Cescutti}, {Chapuis}, {Chiappini}, {Childress}, {Christlieb},
  {Church}, {Cioni}, {Cluver}, {Colless}, {Collett}, {Comparat}, {Cooper},
  {Couch}, {Courbin}, {Croom}, {Croton}, {Daguis{\'e}}, {Dalton}, {Davies},
  {Davis}, {de Laverny}, {Deason}, {Dionies}, {Disseau}, {Doel}, {D{\"o}scher},
  {Driver}, {Dwelly}, {Eckert}, {Edge}, {Edvardsson}, {Youssoufi}, {Elhaddad},
  {Enke}, {Erfanianfar}, {Farrell}, {Fechner}, {Feiz}, {Feltzing}, {Ferreras},
  {Feuerstein}, {Feuillet}, {Finoguenov}, {Ford}, {Fotopoulou}, {Fouesneau},
  {Frenk}, {Frey}, {Gaessler}, {Geier}, {Fusillo}, {Gerhard}, {Giannantonio},
  {Giannone}, {Gibson}, {Gillingham}, {Gonz{\'a}lez-Fern{\'a}ndez},
  {Gonzalez-Solares}, {Gottloeber}, {Gould}, {Grebel}, {Gueguen}, {Guiglion},
  {Haehnelt}, {Hahn}, {Hansen}, {Hartman}, {Hauptner}, {Hawkins}, {Haynes},
  {Haynes}, {Heiter}, {Helmi}, {Aguayo}, {Hewett}, {Hinton}, {Hobbs}, {Hoenig},
  {Hofman}, {Hook}, {Hopgood}, {Hopkins}, {Hourihane}, {Howes}, {Howlett},
  {Huet}, {Irwin}, {Iwert}, {Jablonka}, {Jahn}, {Jahnke}, {Jarno}, {Jin},
  {Jofre}, {Johl}, {Jones}, {J{\"o}nsson}, {Jordan}, {Karovicova}, {Khalatyan},
  {Kelz}, {Kennicutt}, {King}, {Kitaura}, {Klar}, {Klauser}, {Kneib}, {Koch},
  {Koposov}, {Kordopatis}, {Korn}, {Kosmalski}, {Kotak}, {Kovalev}, {Kreckel},
  {Kripak}, {Krumpe}, {Kuijken}, {Kunder}, {Kushniruk}, {Lam}, {Lamer},
  {Laurent}, {Lawrence}, {Lehmitz}, {Lemasle}, {Lewis}, {Li}, {Lidman}, {Lind},
  {Liske}, {Lizon}, {Loveday}, {Ludwig}, {McDermid}, {Maguire}, {Mainieri},
  {Mali}, {Mandel}, {Mandel}, {Mannering}, {Martell}, {Martinez Delgado},
  {Matijevic}, {McGregor}, {McMahon}, {McMillan}, {Mena}, {Merloni}, {Meyer},
  {Michel}, {Micheva}, {Migniau}, {Minchev}, {Monari}, {Muller}, {Murphy},
  {Muthukrishna}, {Nandra}, {Navarro}, {Ness}, {Nichani}, {Nichol}, {Nicklas},
  {Niederhofer}, {Norberg}, {Obreschkow}, {Oliver}, {Owers}, {Pai},
  {Pankratow}, {Parkinson}, {Paschke}, {Paterson}, {Pecontal}, {Parry},
  {Phillips}, {Pillepich}, {Pinard}, {Pirard}, {Piskunov}, {Plank},
  {Pl{\"u}schke}, {Pons}, {Popesso}, {Power}, {Pragt}, {Pramskiy}, {Pryer},
  {Quattri}, {Queiroz}, {Quirrenbach}, {Rahurkar}, {Raichoor}, {Ramstedt},
  {Rau}, {Recio-Blanco}, {Reiss}, {Renaud}, {Revaz}, {Rhode}, {Richard},
  {Richter}, {Rix}, {Robotham}, {Roelfsema}, {Romaniello}, {Rosario},
  {Rothmaier}, {Roukema}, {Ruchti}, {Rupprecht}, {Rybizki}, {Ryde}, {Saar},
  {Sadler}, {Sahl{\'e}n}, {Salvato}, {Sassolas}, {Saunders}, {Saviauk},
  {Sbordone}, {Schmidt}, {Schnurr}, {Scholz}, {Schwope}, {Seifert}, {Shanks},
  {Sheinis}, {Sivov}, {Sk{\'u}lad{\'o}ttir}, {Smartt}, {Smedley}, {Smith},
  {Smith}, {Sorce}, {Spitler}, {Starkenburg}, {Steinmetz}, {Stilz}, {Storm},
  {Sullivan}, {Sutherland}, {Swann}, {Tamone}, {Taylor}, {Teillon}, {Tempel},
  {ter Horst}, {Thi}, {Tolstoy}, {Trager}, {Traven}, {Tremblay}, {Tresse},
  {Valentini}, {van de Weygaert}, {van den Ancker}, {Veljanoski}, {Venkatesan},
  {Wagner}, {Wagner}, {Walcher}, {Waller}, {Walton}, {Wang}, {Winkler},
  {Wisotzki}, {Worley}, {Worseck}, {Xiang}, {Xu}, {Yong}, {Zhao}, {Zheng},
  {Zscheyge}, \& {Zucker}}]{dejong2019}
{de Jong}, R.~S., {Agertz}, O., {Berbel}, A.~A., {et~al.} 2019, The Messenger,
  175, 3

\bibitem[{{Di Teodoro} {et~al.}(2019){Di Teodoro}, {McClure-Griffiths},
  {Jameson}, {D{\'e}nes}, {Dickey}, {Stanimirovi{\'c}}, {Staveley-Smith},
  {Anderson}, {Bunton}, {Chippendale}, {Lee-Waddell}, {MacLeod}, \&
  {Voronkov}}]{diteodoro2019}
{Di Teodoro}, E.~M., {McClure-Griffiths}, N.~M., {Jameson}, K.~E., {et~al.}
  2019, \mnras, 483, 392

\bibitem[{{Diaz} \& {Bekki}(2012)}]{diaz2012}
{Diaz}, J.~D. \& {Bekki}, K. 2012, \apj, 750, 36

\bibitem[{{El Youssoufi} {et~al.}(2021){El Youssoufi}, {Cioni}, {Bell}, {de
  Grijs}, {Groenewegen}, {Ivanov}, {Matijev{\u{i}}c}, {Niederhofer},
  {Oliveira}, {Ripepi}, {Schmidt}, {Subramanian}, {Sun}, \& {van
  Loon}}]{ElYoussoufi2021}
{El Youssoufi}, D., {Cioni}, M.-R.~L., {Bell}, C. P.~M., {et~al.} 2021, \mnras,
  505, 2020

\bibitem[{{El Youssoufi} {et~al.}(2019){El Youssoufi}, {Cioni}, {Bell},
  {Rubele}, {Bekki}, {de Grijs}, {Girardi}, {Ivanov}, {Matijevic},
  {Niederhofer}, {Oliveira}, {Ripepi}, {Subramanian}, \& {van
  Loon}}]{elyoussoufi2019}
{El Youssoufi}, D., {Cioni}, M.-R.~L., {Bell}, C. P.~M., {et~al.} 2019, \mnras,
  490, 1076

\bibitem[{{Emerson} {et~al.}(2006){Emerson}, {Irwin}, \&
  {Hambly}}]{Emerson2006}
{Emerson}, J., {Irwin}, M., \& {Hambly}, N. 2006, in \procspie, Vol. 6270,
  62700S

\bibitem[{{Erkal} {et~al.}(2019){Erkal}, {Belokurov}, {Laporte}, {Koposov},
  {Li}, {Grillmair}, {Kallivayalil}, {Price-Whelan}, {Evans}, {Hawkins},
  {Hendel}, {Mateu}, {Navarro}, {del Pino}, {Slater}, {Sohn}, \& {Orphan Aspen
  Treasury Collaboration}}]{Erkal2019}
{Erkal}, D., {Belokurov}, V., {Laporte}, C.~F.~P., {et~al.} 2019, \mnras, 487,
  2685

\bibitem[{{Gaia Collaboration} {et~al.}(2021{\natexlab{a}}){Gaia
  Collaboration}, {Brown}, {Vallenari}, {Prusti}, {de Bruijne}, {Babusiaux},
  {Biermann}, {Creevey}, {Evans}, {Eyer}, {Hutton}, {Jansen}, {Jordi},
  {Klioner}, {Lammers}, {Lindegren}, {Luri}, {Mignard}, {Panem}, {Pourbaix},
  {Randich}, {Sartoretti}, {Soubiran}, {Walton}, {Arenou}, {Bailer-Jones},
  {Bastian}, {Cropper}, {Drimmel}, {Katz}, {Lattanzi}, {van Leeuwen}, {Bakker},
  {Cacciari}, {Casta{\~n}eda}, {De Angeli}, {Ducourant}, {Fabricius},
  {Fouesneau}, {Fr{\'e}mat}, {Guerra}, {Guerrier}, {Guiraud}, {Jean-Antoine
  Piccolo}, {Masana}, {Messineo}, {Mowlavi}, {Nicolas}, {Nienartowicz},
  {Pailler}, {Panuzzo}, {Riclet}, {Roux}, {Seabroke}, {Sordo}, {Tanga},
  {Th{\'e}venin}, {Gracia-Abril}, {Portell}, {Teyssier}, {Altmann}, {Andrae},
  {Bellas-Velidis}, {Benson}, {Berthier}, {Blomme}, {Brugaletta}, {Burgess},
  {Busso}, {Carry}, {Cellino}, {Cheek}, {Clementini}, {Damerdji}, {Davidson},
  {Delchambre}, {Dell'Oro}, {Fern{\'a}ndez-Hern{\'a}ndez}, {Galluccio},
  {Garc{\'\i}a-Lario}, {Garcia-Reinaldos}, {Gonz{\'a}lez-N{\'u}{\~n}ez},
  {Gosset}, {Haigron}, {Halbwachs}, {Hambly}, {Harrison}, {Hatzidimitriou},
  {Heiter}, {Hern{\'a}ndez}, {Hestroffer}, {Hodgkin}, {Holl}, {Jan{\ss}en},
  {Jevardat de Fombelle}, {Jordan}, {Krone-Martins}, {Lanzafame},
  {L{\"o}ffler}, {Lorca}, {Manteiga}, {Marchal}, {Marrese}, {Moitinho}, {Mora},
  {Muinonen}, {Osborne}, {Pancino}, {Pauwels}, {Petit}, {Recio-Blanco},
  {Richards}, {Riello}, {Rimoldini}, {Robin}, {Roegiers}, {Rybizki}, {Sarro},
  {Siopis}, {Smith}, {Sozzetti}, {Ulla}, {Utrilla}, {van Leeuwen}, {van
  Reeven}, {Abbas}, {Abreu Aramburu}, {Accart}, {Aerts}, {Aguado}, {Ajaj},
  {Altavilla}, {{\'A}lvarez}, {{\'A}lvarez Cid-Fuentes}, {Alves}, {Anderson},
  {Anglada Varela}, {Antoja}, {Audard}, {Baines}, {Baker},
  {Balaguer-N{\'u}{\~n}ez}, {Balbinot}, {Balog}, {Barache}, {Barbato},
  {Barros}, {Barstow}, {Bartolom{\'e}}, {Bassilana}, {Bauchet},
  {Baudesson-Stella}, {Becciani}, {Bellazzini}, {Bernet}, {Bertone}, {Bianchi},
  {Blanco-Cuaresma}, {Boch}, {Bombrun}, {Bossini}, {Bouquillon}, {Bragaglia},
  {Bramante}, {Breedt}, {Bressan}, {Brouillet}, {Bucciarelli}, {Burlacu},
  {Busonero}, {Butkevich}, {Buzzi}, {Caffau}, {Cancelliere}, {C{\'a}novas},
  {Cantat-Gaudin}, {Carballo}, {Carlucci}, {Carnerero}, {Carrasco},
  {Casamiquela}, {Castellani}, {Castro-Ginard}, {Castro Sampol}, {Chaoul},
  {Charlot}, {Chemin}, {Chiavassa}, {Cioni}, {Comoretto}, {Cooper}, {Cornez},
  {Cowell}, {Crifo}, {Crosta}, {Crowley}, {Dafonte}, {Dapergolas}, {David},
  {David}, {de Laverny}, {De Luise}, {De March}, {De Ridder}, {de Souza}, {de
  Teodoro}, {de Torres}, {del Peloso}, {del Pozo}, {Delbo}, {Delgado},
  {Delgado}, {Delisle}, {Di Matteo}, {Diakite}, {Diener}, {Distefano},
  {Dolding}, {Eappachen}, {Edvardsson}, {Enke}, {Esquej}, {Fabre}, {Fabrizio},
  {Faigler}, {Fedorets}, {Fernique}, {Fienga}, {Figueras}, {Fouron},
  {Fragkoudi}, {Fraile}, {Franke}, {Gai}, {Garabato}, {Garcia-Gutierrez},
  {Garc{\'\i}a-Torres}, {Garofalo}, {Gavras}, {Gerlach}, {Geyer}, {Giacobbe},
  {Gilmore}, {Girona}, {Giuffrida}, {Gomel}, {Gomez}, {Gonzalez-Santamaria},
  {Gonz{\'a}lez-Vidal}, {Granvik}, {Guti{\'e}rrez-S{\'a}nchez}, {Guy},
  {Hauser}, {Haywood}, {Helmi}, {Hidalgo}, {Hilger}, {H{\l}adczuk}, {Hobbs},
  {Holland}, {Huckle}, {Jasniewicz}, {Jonker}, {Juaristi Campillo}, {Julbe},
  {Karbevska}, {Kervella}, {Khanna}, {Kochoska}, {Kontizas}, {Kordopatis},
  {Korn}, {Kostrzewa-Rutkowska}, {Kruszy{\'n}ska}, {Lambert}, {Lanza}, {Lasne},
  {Le Campion}, {Le Fustec}, {Lebreton}, {Lebzelter}, {Leccia}, {Leclerc},
  {Lecoeur-Taibi}, {Liao}, {Licata}, {Lindstr{\o}m}, {Lister}, {Livanou},
  {Lobel}, {Madrero Pardo}, {Managau}, {Mann}, {Marchant}, {Marconi}, {Marcos
  Santos}, {Marinoni}, {Marocco}, {Marshall}, {Martin Polo},
  {Mart{\'\i}n-Fleitas}, {Masip}, {Massari}, {Mastrobuono-Battisti}, {Mazeh},
  {McMillan}, {Messina}, {Michalik}, {Millar}, {Mints}, {Molina}, {Molinaro},
  {Moln{\'a}r}, {Montegriffo}, {Mor}, {Morbidelli}, {Morel}, {Morris},
  {Mulone}, {Munoz}, {Muraveva}, {Murphy}, {Musella}, {Noval}, {Ord{\'e}novic},
  {Orr{\`u}}, {Osinde}, {Pagani}, {Pagano}, {Palaversa}, {Palicio}, {Panahi},
  {Pawlak}, {Pe{\~n}alosa Esteller}, {Penttil{\"a}}, {Piersimoni}, {Pineau},
  {Plachy}, {Plum}, {Poggio}, {Poretti}, {Poujoulet}, {Pr{\v{s}}a}, {Pulone},
  {Racero}, {Ragaini}, {Rainer}, {Raiteri}, {Rambaux}, {Ramos}, {Ramos-Lerate},
  {Re Fiorentin}, {Regibo}, {Reyl{\'e}}, {Ripepi}, {Riva}, {Rixon}, {Robichon},
  {Robin}, {Roelens}, {Rohrbasser}, {Romero-G{\'o}mez}, {Rowell}, {Royer},
  {Rybicki}, {Sadowski}, {Sagrist{\`a} Sell{\'e}s}, {Sahlmann}, {Salgado},
  {Salguero}, {Samaras}, {Sanchez Gimenez}, {Sanna}, {Santove{\~n}a},
  {Sarasso}, {Schultheis}, {Sciacca}, {Segol}, {Segovia}, {S{\'e}gransan},
  {Semeux}, {Shahaf}, {Siddiqui}, {Siebert}, {Siltala}, {Slezak}, {Smart},
  {Solano}, {Solitro}, {Souami}, {Souchay}, {Spagna}, {Spoto}, {Steele},
  {Steidelm{\"u}ller}, {Stephenson}, {S{\"u}veges}, {Szabados}, {Szegedi-Elek},
  {Taris}, {Tauran}, {Taylor}, {Teixeira}, {Thuillot}, {Tonello}, {Torra},
  {Torra}, {Turon}, {Unger}, {Vaillant}, {van Dillen}, {Vanel}, {Vecchiato},
  {Viala}, {Vicente}, {Voutsinas}, {Weiler}, {Wevers}, {Wyrzykowski}, {Yoldas},
  {Yvard}, {Zhao}, {Zorec}, {Zucker}, {Zurbach}, \& {Zwitter}}]{brown2020}
{Gaia Collaboration}, {Brown}, A.~G.~A., {Vallenari}, A., {et~al.}
  2021{\natexlab{a}}, \aap, 650, C3

\bibitem[{{Gaia Collaboration} {et~al.}(2018){Gaia Collaboration}, {Helmi},
  {van Leeuwen}, {McMillan}, {Massari}, {Antoja}, {Robin}, {Lindegren},
  {Bastian}, {Arenou}, {Babusiaux}, {Biermann}, {Breddels}, {Hobbs}, {Jordi},
  {Pancino}, {Reyl{\'e}}, {Veljanoski}, {Brown}, {Vallenari}, {Prusti}, {de
  Bruijne}, {Bailer-Jones}, {Evans}, {Eyer}, {Jansen}, {Klioner}, {Lammers},
  {Luri}, {Mignard}, {Panem}, {Pourbaix}, {Randich}, {Sartoretti}, {Siddiqui},
  {Soubiran}, {Walton}, {Cropper}, {Drimmel}, {Katz}, {Lattanzi}, {Bakker},
  {Cacciari}, {Casta{\~n}eda}, {Chaoul}, {Cheek}, {De Angeli}, {Fabricius},
  {Guerra}, {Holl}, {Masana}, {Messineo}, {Mowlavi}, {Nienartowicz}, {Panuzzo},
  {Portell}, {Riello}, {Seabroke}, {Tanga}, {Th{\'e}venin}, {Gracia-Abril},
  {Comoretto}, {Garcia-Reinaldos}, {Teyssier}, {Altmann}, {Andrae}, {Audard},
  {Bellas-Velidis}, {Benson}, {Berthier}, {Blomme}, {Burgess}, {Busso},
  {Carry}, {Cellino}, {Clementini}, {Clotet}, {Creevey}, {Davidson}, {De
  Ridder}, {Delchambre}, {Dell'Oro}, {Ducourant},
  {Fern{\'a}ndez-Hern{\'a}ndez}, {Fouesneau}, {Fr{\'e}mat}, {Galluccio},
  {Garc{\'\i}a-Torres}, {Gonz{\'a}lez-N{\'u}{\~n}ez}, {Gonz{\'a}lez-Vidal},
  {Gosset}, {Guy}, {Halbwachs}, {Hambly}, {Harrison}, {Hern{\'a}ndez},
  {Hestroffer}, {Hodgkin}, {Hutton}, {Jasniewicz}, {Jean-Antoine-Piccolo},
  {Jordan}, {Korn}, {Krone-Martins}, {Lanzafame}, {Lebzelter}, {L{\"o}ffler},
  {Manteiga}, {Marrese}, {Mart{\'\i}n-Fleitas}, {Moitinho}, {Mora}, {Muinonen},
  {Osinde}, {Pauwels}, {Petit}, {Recio-Blanco}, {Richards}, {Rimoldini},
  {Sarro}, {Siopis}, {Smith}, {Sozzetti}, {S{\"u}veges}, {Torra}, {van Reeven},
  {Abbas}, {Abreu Aramburu}, {Accart}, {Aerts}, {Altavilla}, {{\'A}lvarez},
  {Alvarez}, {Alves}, {Anderson}, {Andrei}, {Anglada Varela}, {Antiche},
  {Arcay}, {Astraatmadja}, {Bach}, {Baker}, {Balaguer-N{\'u}{\~n}ez}, {Balm},
  {Barache}, {Barata}, {Barbato}, {Barblan}, {Barklem}, {Barrado}, {Barros},
  {Barstow}, {Bartholom{\'e} Mu{\~n}oz}, {Bassilana}, {Becciani}, {Bellazzini},
  {Berihuete}, {Bertone}, {Bianchi}, {Bienaym{\'e}}, {Blanco-Cuaresma}, {Boch},
  {Boeche}, {Bombrun}, {Borrachero}, {Bossini}, {Bouquillon}, {Bourda},
  {Bragaglia}, {Bramante}, {Bressan}, {Brouillet}, {Br{\"u}semeister},
  {Brugaletta}, {Bucciarelli}, {Burlacu}, {Busonero}, {Butkevich}, {Buzzi},
  {Caffau}, {Cancelliere}, {Cannizzaro}, {Cantat-Gaudin}, {Carballo},
  {Carlucci}, {Carrasco}, {Casamiquela}, {Castellani}, {Castro-Ginard},
  {Charlot}, {Chemin}, {Chiavassa}, {Cocozza}, {Costigan}, {Cowell}, {Crifo},
  {Crosta}, {Crowley}, {Cuypers}, {Dafonte}, {Damerdji}, {Dapergolas}, {David},
  {David}, {de Laverny}, {De Luise}, {De March}, {de Martino}, {de Souza}, {de
  Torres}, {Debosscher}, {del Pozo}, {Delbo}, {Delgado}, {Delgado}, {Di
  Matteo}, {Diakite}, {Diener}, {Distefano}, {Dolding}, {Drazinos},
  {Dur{\'a}n}, {Edvardsson}, {Enke}, {Eriksson}, {Esquej}, {Eynard Bontemps},
  {Fabre}, {Fabrizio}, {Faigler}, {Falc{\~a}o}, {Farr{\`a}s Casas}, {Federici},
  {Fedorets}, {Fernique}, {Figueras}, {Filippi}, {Findeisen}, {Fonti},
  {Fraile}, {Fraser}, {Fr{\'e}zouls}, {Gai}, {Galleti}, {Garabato},
  {Garc{\'\i}a-Sedano}, {Garofalo}, {Garralda}, {Gavel}, {Gavras}, {Gerssen},
  {Geyer}, {Giacobbe}, {Gilmore}, {Girona}, {Giuffrida}, {Glass}, {Gomes},
  {Granvik}, {Gueguen}, {Guerrier}, {Guiraud}, {Guti{\'e}rrez-S{\'a}nchez},
  {Hofmann}, {Holland}, {Huckle}, {Hypki}, {Icardi}, {Jan{\ss}en}, {Jevardat de
  Fombelle}, {Jonker}, {Juh{\'a}sz}, {Julbe}, {Karampelas}, {Kewley}, {Klar},
  {Kochoska}, {Kohley}, {Kolenberg}, {Kontizas}, {Kontizas}, {Koposov},
  {Kordopatis}, {Kostrzewa-Rutkowska}, {Koubsky}, {Lambert}, {Lanza}, {Lasne},
  {Lavigne}, {Le Fustec}, {Le Poncin-Lafitte}, {Lebreton}, {Leccia}, {Leclerc},
  {Lecoeur-Taibi}, {Lenhardt}, {Leroux}, {Liao}, {Licata}, {Lindstr{\o}m},
  {Lister}, {Livanou}, {Lobel}, {L{\'o}pez}, {Managau}, {Mann}, {Mantelet},
  {Marchal}, {Marchant}, {Marconi}, {Marinoni}, {Marschalk{\'o}}, {Marshall},
  {Martino}, {Marton}, {Mary}, {Matijevi{\v{c}}}, {Mazeh}, {Messina},
  {Michalik}, {Millar}, {Molina}, {Molinaro}, {Moln{\'a}r}, {Montegriffo},
  {Mor}, {Morbidelli}, {Morel}, {Morris}, {Mulone}, {Muraveva}, {Musella},
  {Nelemans}, {Nicastro}, {Noval}, {O'Mullane}, {Ord{\'e}novic},
  {Ord{\'o}{\~n}ez-Blanco}, {Osborne}, {Pagani}, {Pagano}, {Pailler},
  {Palacin}, {Palaversa}, {Panahi}, {Pawlak}, {Piersimoni}, {Pineau}, {Plachy},
  {Plum}, {Poggio}, {Poujoulet}, {Pr{\v{s}}a}, {Pulone}, {Racero}, {Ragaini},
  {Rambaux}, {Ramos-Lerate}, {Regibo}, {Riclet}, {Ripepi}, {Riva}, {Rivard},
  {Rixon}, {Roegiers}, {Roelens}, {Romero-G{\'o}mez}, {Rowell}, {Royer},
  {Ruiz-Dern}, {Sadowski}, {Sagrist{\`a} Sell{\'e}s}, {Sahlmann}, {Salgado},
  {Salguero}, {Sanna}, {Santana-Ros}, {Sarasso}, {Savietto}, {Schultheis},
  {Sciacca}, {Segol}, {Segovia}, {S{\'e}gransan}, {Shih}, {Siltala}, {Silva},
  {Smart}, {Smith}, {Solano}, {Solitro}, {Sordo}, {Soria Nieto}, {Souchay},
  {Spagna}, {Spoto}, {Stampa}, {Steele}, {Steidelm{\"u}ller}, {Stephenson},
  {Stoev}, {Suess}, {Surdej}, {Szabados}, {Szegedi-Elek}, {Tapiador}, {Taris},
  {Tauran}, {Taylor}, {Teixeira}, {Terrett}, {Teyssand ier}, {Thuillot},
  {Titarenko}, {Torra Clotet}, {Turon}, {Ulla}, {Utrilla}, {Uzzi}, {Vaillant},
  {Valentini}, {Valette}, {van Elteren}, {Van Hemelryck}, {van Leeuwen},
  {Vaschetto}, {Vecchiato}, {Viala}, {Vicente}, {Vogt}, {von Essen}, {Voss},
  {Votruba}, {Voutsinas}, {Walmsley}, {Weiler}, {Wertz}, {Wevems},
  {Wyrzykowski}, {Yoldas}, {{\v{Z}}erjal}, {Ziaeepour}, {Zorec}, {Zschocke},
  {Zucker}, {Zurbach}, \& {Zwitter}}]{helmi2018}
{Gaia Collaboration}, {Helmi}, A., {van Leeuwen}, F., {et~al.} 2018, \aap, 616,
  A12

\bibitem[{{Gaia Collaboration} {et~al.}(2021{\natexlab{b}}){Gaia
  Collaboration}, {Luri}, {Chemin}, {Clementini}, {Delgado}, {McMillan},
  {Romero-G{\'o}mez}, {Balbinot}, {Castro-Ginard}, {Mor}, {Ripepi}, {Sarro},
  {Cioni}, {Fabricius}, {Garofalo}, {Helmi}, {Muraveva}, {Brown}, {Vallenari},
  {Prusti}, {de Bruijne}, {Babusiaux}, {Biermann}, {Creevey}, {Evans}, {Eyer},
  {Hutton}, {Jansen}, {Jordi}, {Klioner}, {Lammers}, {Lindegren}, {Mignard},
  {Panem}, {Pourbaix}, {Randich}, {Sartoretti}, {Soubiran}, {Walton}, {Arenou},
  {Bailer-Jones}, {Bastian}, {Cropper}, {Drimmel}, {Katz}, {Lattanzi}, {van
  Leeuwen}, {Bakker}, {Casta{\~n}eda}, {De Angeli}, {Ducourant}, {Fouesneau},
  {Fr{\'e}mat}, {Guerra}, {Guerrier}, {Guiraud}, {Jean-Antoine Piccolo},
  {Masana}, {Messineo}, {Mowlavi}, {Nicolas}, {Nienartowicz}, {Pailler},
  {Panuzzo}, {Riclet}, {Roux}, {Seabroke}, {Sordo}, {Tanga}, {Th{\'e}venin},
  {Gracia-Abril}, {Portell}, {Teyssier}, {Altmann}, {Andrae}, {Bellas-Velidis},
  {Benson}, {Berthier}, {Blomme}, {Brugaletta}, {Burgess}, {Busso}, {Carry},
  {Cellino}, {Cheek}, {Damerdji}, {Davidson}, {Delchambre}, {Dell'Oro},
  {Fern{\'a}ndez-Hern{\'a}ndez}, {Galluccio}, {Garc{\'\i}a-Lario},
  {Garcia-Reinaldos}, {Gonz{\'a}lez-N{\'u}{\~n}ez}, {Gosset}, {Haigron},
  {Halbwachs}, {Hambly}, {Harrison}, {Hatzidimitriou}, {Heiter},
  {Hern{\'a}ndez}, {Hestroffer}, {Hodgkin}, {Holl}, {Jan{\ss}en}, {Jevardat de
  Fombelle}, {Jordan}, {Krone-Martins}, {Lanzafame}, {L{\"o}ffler}, {Lorca},
  {Manteiga}, {Marchal}, {Marrese}, {Moitinho}, {Mora}, {Muinonen}, {Osborne},
  {Pancino}, {Pauwels}, {Recio-Blanco}, {Richards}, {Riello}, {Rimoldini},
  {Robin}, {Roegiers}, {Rybizki}, {Siopis}, {Smith}, {Sozzetti}, {Ulla},
  {Utrilla}, {van Leeuwen}, {van Reeven}, {Abbas}, {Abreu Aramburu}, {Accart},
  {Aerts}, {Aguado}, {Ajaj}, {Altavilla}, {{\'A}lvarez}, {{\'A}lvarez
  Cid-Fuentes}, {Alves}, {Anderson}, {Anglada Varela}, {Antoja}, {Audard},
  {Baines}, {Baker}, {Balaguer-N{\'u}{\~n}ez}, {Balog}, {Barache}, {Barbato},
  {Barros}, {Barstow}, {Bartolom{\'e}}, {Bassilana}, {Bauchet},
  {Baudesson-Stella}, {Becciani}, {Bellazzini}, {Bernet}, {Bertone}, {Bianchi},
  {Blanco-Cuaresma}, {Boch}, {Bombrun}, {Bossini}, {Bouquillon}, {Bragaglia},
  {Bramante}, {Breedt}, {Bressan}, {Brouillet}, {Bucciarelli}, {Burlacu},
  {Busonero}, {Butkevich}, {Buzzi}, {Caffau}, {Cancelliere}, {C{\'a}novas},
  {Cantat-Gaudin}, {Carballo}, {Carlucci}, {Carnerero}, {Carrasco},
  {Casamiquela}, {Castellani}, {Castro Sampol}, {Chaoul}, {Charlot},
  {Chiavassa}, {Comoretto}, {Cooper}, {Cornez}, {Cowell}, {Crifo}, {Crosta},
  {Crowley}, {Dafonte}, {Dapergolas}, {David}, {David}, {de Laverny}, {De
  Luise}, {De March}, {De Ridder}, {de Souza}, {de Teodoro}, {de Torres}, {del
  Peloso}, {del Pozo}, {Delgado}, {Delisle}, {Di Matteo}, {Diakite}, {Diener},
  {Distefano}, {Dolding}, {Eappachen}, {Enke}, {Esquej}, {Fabre}, {Fabrizio},
  {Faigler}, {Fedorets}, {Fernique}, {Fienga}, {Figueras}, {Fouron},
  {Fragkoudi}, {Fraile}, {Franke}, {Gai}, {Garabato}, {Garcia-Gutierrez},
  {Garc{\'\i}a-Torres}, {Gavras}, {Gerlach}, {Geyer}, {Giacobbe}, {Gilmore},
  {Girona}, {Giuffrida}, {Gomez}, {Gonzalez-Santamaria}, {Gonz{\'a}lez-Vidal},
  {Granvik}, {Guti{\'e}rrez-S{\'a}nchez}, {Guy}, {Hauser}, {Haywood},
  {Hidalgo}, {Hilger}, {H{\l}adczuk}, {Hobbs}, {Holland}, {Huckle},
  {Jasniewicz}, {Jonker}, {Juaristi Campillo}, {Julbe}, {Karbevska},
  {Kervella}, {Khanna}, {Kochoska}, {Kontizas}, {Kordopatis}, {Korn},
  {Kostrzewa-Rutkowska}, {Kruszy{\'n}ska}, {Lambert}, {Lanza}, {Lasne}, {Le
  Campion}, {Le Fustec}, {Lebreton}, {Lebzelter}, {Leccia}, {Leclerc},
  {Lecoeur-Taibi}, {Liao}, {Licata}, {Lindstr{\o}m}, {Lister}, {Livanou},
  {Lobel}, {Madrero Pardo}, {Managau}, {Mann}, {Marchant}, {Marconi}, {Marcos
  Santos}, {Marinoni}, {Marocco}, {Marshall}, {Martin Polo},
  {Mart{\'\i}n-Fleitas}, {Masip}, {Massari}, {Mastrobuono-Battisti}, {Mazeh},
  {Messina}, {Michalik}, {Millar}, {Mints}, {Molina}, {Molinaro}, {Moln{\'a}r},
  {Montegriffo}, {Morbidelli}, {Morel}, {Morris}, {Mulone}, {Munoz}, {Murphy},
  {Musella}, {Noval}, {Ord{\'e}novic}, {Orr{\`u}}, {Osinde}, {Pagani},
  {Pagano}, {Palaversa}, {Palicio}, {Panahi}, {Pawlak}, {Pe{\~n}alosa
  Esteller}, {Penttil{\"a}}, {Piersimoni}, {Pineau}, {Plachy}, {Plum},
  {Poggio}, {Poretti}, {Poujoulet}, {Pr{\v{s}}a}, {Pulone}, {Racero},
  {Ragaini}, {Rainer}, {Raiteri}, {Rambaux}, {Ramos}, {Ramos-Lerate}, {Re
  Fiorentin}, {Regibo}, {Reyl{\'e}}, {Riva}, {Rixon}, {Robichon}, {Robin},
  {Roelens}, {Rohrbasser}, {Rowell}, {Royer}, {Rybicki}, {Sadowski},
  {Sagrist{\`a} Sell{\'e}s}, {Sahlmann}, {Salgado}, {Salguero}, {Samaras},
  {Gimenez}, {Sanna}, {Santove{\~n}a}, {Sarasso}, {Schultheis}, {Sciacca},
  {Segol}, {Segovia}, {S{\'e}gransan}, {Semeux}, {Siddiqui}, {Siebert},
  {Siltala}, {Slezak}, {Smart}, {Solano}, {Solitro}, {Souami}, {Souchay},
  {Spagna}, {Spoto}, {Steele}, {Steidelm{\"u}ller}, {Stephenson},
  {S{\"u}veges}, {Szabados}, {Szegedi-Elek}, {Taris}, {Tauran}, {Taylor},
  {Teixeira}, {Thuillot}, {Tonello}, {Torra}, {Torra}, {Turon}, {Unger},
  {Vaillant}, {van Dillen}, {Vanel}, {Vecchiato}, {Viala}, {Vicente},
  {Voutsinas}, {Weiler}, {Wevers}, {Wyrzykowski}, {Yoldas}, {Yvard}, {Zhao},
  {Zorec}, {Zucker}, {Zurbach}, \& {Zwitter}}]{luri2020}
{Gaia Collaboration}, {Luri}, X., {Chemin}, L., {et~al.} 2021{\natexlab{b}},
  \aap, 649, A7

\bibitem[{{Gardiner} \& {Noguchi}(1996)}]{gardiner1996}
{Gardiner}, L.~T. \& {Noguchi}, M. 1996, JKAS Suppl., 29, S93

\bibitem[{{Gonz{\'a}lez-Fern{\'a}ndez}
  {et~al.}(2018){Gonz{\'a}lez-Fern{\'a}ndez}, {Hodgkin}, {Irwin},
  {Gonz{\'a}lez-Solares}, {Koposov}, {Lewis}, {Emerson}, {Hewett},
  {Yolda{\textcommabelow s}}, \& {Riello}}]{gonzalez2018}
{Gonz{\'a}lez-Fern{\'a}ndez}, C., {Hodgkin}, S.~T., {Irwin}, M.~J., {et~al.}
  2018, \mnras, 474, 5459

\bibitem[{{Hammer} {et~al.}(2015){Hammer}, {Yang}, {Flores}, {Puech}, \&
  {Fouquet}}]{Hammer2015}
{Hammer}, F., {Yang}, Y.~B., {Flores}, H., {Puech}, M., \& {Fouquet}, S. 2015,
  \apj, 813, 110

\bibitem[{{Hashimoto} {et~al.}(2003){Hashimoto}, {Funato}, \&
  {Makino}}]{Hashimoto2003}
{Hashimoto}, Y., {Funato}, Y., \& {Makino}, J. 2003, \apj, 582, 196

\bibitem[{{Indu} \& {Subramaniam}(2015)}]{indu2015}
{Indu}, G. \& {Subramaniam}, A. 2015, \aap, 573, A136

\bibitem[{{Kruijssen} {et~al.}(2020){Kruijssen}, {Pfeffer}, {Chevance},
  {Bonaca}, {Trujillo-Gomez}, {Bastian}, {Reina-Campos}, {Crain}, \&
  {Hughes}}]{Kruijssen2020}
{Kruijssen}, J.~M.~D., {Pfeffer}, J.~L., {Chevance}, M., {et~al.} 2020, \mnras,
  498, 2472

\bibitem[{{Lindegren} {et~al.}(2021){Lindegren}, {Klioner}, {Hern{\'a}ndez},
  {Bombrun}, {Ramos-Lerate}, {Steidelm{\"u}ller}, {Bastian}, {Biermann}, {de
  Torres}, {Gerlach}, {Geyer}, {Hilger}, {Hobbs}, {Lammers}, {McMillan},
  {Stephenson}, {Casta{\~n}eda}, {Davidson}, {Fabricius}, {Gracia-Abril},
  {Portell}, {Rowell}, {Teyssier}, {Torra}, {Bartolom{\'e}}, {Clotet},
  {Garralda}, {Gonz{\'a}lez-Vidal}, {Torra}, {Abbas}, {Altmann}, {Anglada
  Varela}, {Balaguer-N{\'u}{\~n}ez}, {Balog}, {Barache}, {Becciani}, {Bernet},
  {Bertone}, {Bianchi}, {Bouquillon}, {Brown}, {Bucciarelli}, {Busonero},
  {Butkevich}, {Buzzi}, {Cancelliere}, {Carlucci}, {Charlot}, {Cioni},
  {Crosta}, {Crowley}, {del Peloso}, {del Pozo}, {Drimmel}, {Esquej}, {Fienga},
  {Fraile}, {Gai}, {Garcia-Reinaldos}, {Guerra}, {Hambly}, {Hauser},
  {Jan{\ss}en}, {Jordan}, {Kostrzewa-Rutkowska}, {Lattanzi}, {Liao}, {Licata},
  {Lister}, {L{\"o}ffler}, {Marchant}, {Masip}, {Mignard}, {Mints}, {Molina},
  {Mora}, {Morbidelli}, {Murphy}, {Pagani}, {Panuzzo}, {Pe{\~n}alosa Esteller},
  {Poggio}, {Re Fiorentin}, {Riva}, {Sagrist{\`a} Sell{\'e}s}, {Sanchez
  Gimenez}, {Sarasso}, {Sciacca}, {Siddiqui}, {Smart}, {Souami}, {Spagna},
  {Steele}, {Taris}, {Utrilla}, {van Reeven}, \& {Vecchiato}}]{klioner2020}
{Lindegren}, L., {Klioner}, S.~A., {Hern{\'a}ndez}, J., {et~al.} 2021, \aap,
  649, A2

\bibitem[{{Lopes} \& {Ribeiro}(2020)}]{lopes2020}
{Lopes}, P. A.~A. \& {Ribeiro}, A. L.~B. 2020, \mnras, 493, 3429

\bibitem[{{Martin} {et~al.}(2004){Martin}, {Ibata}, {Bellazzini}, {Irwin},
  {Lewis}, \& {Dehnen}}]{Martin2004}
{Martin}, N.~F., {Ibata}, R.~A., {Bellazzini}, M., {et~al.} 2004, \mnras, 348,
  12

\bibitem[{{Mazzi} {et~al.}(2021){Mazzi}, {Girardi}, {Zaggia}, {Pastorelli},
  {Rubele}, {Bressan}, {Cioni}, {Clementini}, {Cusano}, {Rocha}, {Gullieuszik},
  {Kerber}, {Marigo}, {Ripepi}, {Bekki}, {Bell}, {de Grijs}, {Groenewegen},
  {Ivanov}, {Oliveira}, {Sun}, \& {van Loon}}]{Mazzi2021}
{Mazzi}, A., {Girardi}, L., {Zaggia}, S., {et~al.} 2021, \mnras, 508, 245

\bibitem[{{McClure-Griffiths} {et~al.}(2018){McClure-Griffiths}, {D{\'e}nes},
  {Dickey}, {Stanimirovi{\'c}}, {}, {Staveley-Smith}, {Jameson}, {Di Teodoro},
  {Allison}, {Collier}, {Chippendale}, {Franzen}, {G{\"u}rkan}, {Heald},
  {Hotan}, {Kleiner}, {Lee-Waddell}, {McConnell}, {Popping}, {Rhee}, {Riseley},
  {Voronkov}, \& {Whiting}}]{mcclure-griffiths2018}
{McClure-Griffiths}, N.~M., {D{\'e}nes}, H., {Dickey}, J.~M., {et~al.} 2018,
  Nature Astronomy, 2, 901

\bibitem[{Menon {et~al.}(2012)Menon, Jiang, Vembu, Elkan, \&
  Ohno-Machado}]{Menon2012}
Menon, A., Jiang, X., Vembu, S., Elkan, C., \& Ohno-Machado, L. 2012,
  Proceedings of the 29th International Conference on Machine Learning, ICML
  2012, 1

\bibitem[{{Mucciarelli} {et~al.}(2017){Mucciarelli}, {Bellazzini}, {Ibata},
  {Romano}, {Chapman}, \& {Monaco}}]{Mucciarelli2017}
{Mucciarelli}, A., {Bellazzini}, M., {Ibata}, R., {et~al.} 2017, \aap, 605, A46

\bibitem[{Niculescu-Mizil \& Caruana(2005)}]{Niculescu-Mizil2005}
Niculescu-Mizil, A. \& Caruana, R. 2005, in Proceedings of the 22nd
  International Conference on Machine Learning, ICML '05 (New York, NY, USA:
  Association for Computing Machinery), 625–632

\bibitem[{{Niederhofer} {et~al.}(2018){Niederhofer}, {Cioni}, {Rubele},
  {Schmidt}, {Bekki}, {de Grijs}, {Emerson}, {Ivanov}, {Oliveira},
  {Petr-Gotzens}, {Ripepi}, {Sun}, \& {van Loon}}]{niederhofer2018a}
{Niederhofer}, F., {Cioni}, M.-R.~L., {Rubele}, S., {et~al.} 2018, \aap, 612,
  A115

\bibitem[{{Niederhofer} {et~al.}(2021{\natexlab{a}}){Niederhofer}, {Cioni},
  {Rubele}, {Schmidt}, {Diaz}, {Matijev{\u{i}}c}, {Bekki}, {Bell}, {de Grijs},
  {El Youssoufi}, {Ivanov}, {Oliveira}, {Ripepi}, {Subramanian}, {Sun}, \& {van
  Loon}}]{niederhofer2021}
{Niederhofer}, F., {Cioni}, M.-R.~L., {Rubele}, S., {et~al.}
  2021{\natexlab{a}}, \mnras, 502, 2859

\bibitem[{{Niederhofer} {et~al.}(2021{\natexlab{b}}){Niederhofer}, {Cioni},
  {Schmidt}, {Bekki}, {de Grijs}, {Ivanov}, {Oliveira}, {Ripepi},
  {Subramanian}, \& {van Loon}}]{niederhofer2021b}
{Niederhofer}, F., {Cioni}, M.-R.~L., {Schmidt}, T., {et~al.}
  2021{\natexlab{b}}, \mnras, submitted

\bibitem[{{Olsen} {et~al.}(2015){Olsen}, {Blum}, {Smart}, {Zaritsky}, {Boyer},
  {Gordon}, \& {Massey}}]{olsen2015}
{Olsen}, K.~A.~G., {Blum}, R.~D., {Smart}, B., {et~al.} 2015, in Astronomical
  Society of the Pacific Conference Series, Vol. 491, Fifty Years of Wide Field
  Studies in the Southern Hemisphere: Resolved Stellar Populations of the
  Galactic Bulge and Magellanic Clouds, ed. S.~{Points} \& A.~{Kunder}, 257

\bibitem[{{Olsen} {et~al.}(2011){Olsen}, {Zaritsky}, {Blum}, {Boyer}, \&
  {Gordon}}]{olsen2011}
{Olsen}, K. A.~G., {Zaritsky}, D., {Blum}, R.~D., {Boyer}, M.~L., \& {Gordon},
  K.~D. 2011, \apj, 737, 29

\bibitem[{{Parada} {et~al.}(2021){Parada}, {Heyl}, {Richer}, {Ripoche}, \&
  {Rousseau-Nepton}}]{parada2021}
{Parada}, J., {Heyl}, J., {Richer}, H., {Ripoche}, P., \& {Rousseau-Nepton}, L.
  2021, \mnras, 501, 933

\bibitem[{{Patel} {et~al.}(2017){Patel}, {Besla}, \& {Sohn}}]{Patel2017}
{Patel}, E., {Besla}, G., \& {Sohn}, S.~T. 2017, \mnras, 464, 3825

\bibitem[{{Paturel} {et~al.}(2003){Paturel}, {Petit}, {Prugniel}, {Theureau},
  {Rousseau}, {Brouty}, {Dubois}, \& {Cambr{\'e}sy}}]{paturel2003}
{Paturel}, G., {Petit}, C., {Prugniel}, P., {et~al.} 2003, \aap, 412, 45

\bibitem[{{Pearson} {et~al.}(2018){Pearson}, {Privon}, {Besla}, {Putman},
  {Mart{\'\i}nez-Delgado}, {Johnston}, {Gabany}, {Patton}, \&
  {Kallivayalil}}]{Pearson2018}
{Pearson}, S., {Privon}, G.~C., {Besla}, G., {et~al.} 2018, \mnras, 480, 3069

\bibitem[{{Queiroz} {et~al.}(2018){Queiroz}, {Anders}, {Santiago}, {Chiappini},
  {Steinmetz}, {Dal Ponte}, {Stassun}, {da Costa}, {Maia}, {Crestani}, {Beers},
  {Fern{\'a}ndez-Trincado}, {Garc{\'\i}a-Hern{\'a}ndez}, {Roman-Lopes}, \&
  {Zamora}}]{Queiroz2018}
{Queiroz}, A.~B.~A., {Anders}, F., {Santiago}, B.~X., {et~al.} 2018, \mnras,
  476, 2556

\bibitem[{{Robin} {et~al.}(2012){Robin}, {Luri}, {Reyl{\'e}}, {Isasi}, {Grux},
  {Blanco-Cuaresma}, {Arenou}, {Babusiaux}, {Belcheva}, {Drimmel}, {Jordi},
  {Krone-Martins}, {Masana}, {Mauduit}, {Mignard}, {Mowlavi},
  {Rocca-Volmerange}, {Sartoretti}, {Slezak}, \& {Sozzetti}}]{robin2012}
{Robin}, A.~C., {Luri}, X., {Reyl{\'e}}, C., {et~al.} 2012, \aap, 543, A100

\bibitem[{{Rybizki} {et~al.}(2020){Rybizki}, {Demleitner}, {Bailer-Jones},
  {Tio}, {Cantat-Gaudin}, {Fouesneau}, {Chen}, {Andrae}, {Girardi}, \&
  {Sharma}}]{rybizki2020}
{Rybizki}, J., {Demleitner}, M., {Bailer-Jones}, C., {et~al.} 2020, \pasp, 132,
  074501

\bibitem[{{Schmidt} {et~al.}(2019){Schmidt}, {Cioni}, {Niederhofer}, {Diaz}, \&
  {Matijevic}}]{Schmidt2018}
{Schmidt}, T., {Cioni}, M.-R., {Niederhofer}, F., {Diaz}, J., \& {Matijevic},
  G. 2019, in IAU Symposium, Vol. 344, Dwarf Galaxies: From the Deep Universe
  to the Present, ed. K.~B.~W. {McQuinn} \& S.~{Stierwalt}, 130--133

\bibitem[{{Schmidt} {et~al.}(2020){Schmidt}, {Cioni}, {Niederhofer}, {Bekki},
  {Bell}, {de Grijs}, {Diaz}, {El Youssoufi}, {Emerson}, {Groenewegen},
  {Ivanov}, {Matijevic}, {Oliveira}, {Petr-Gotzens}, {Queiroz}, {Ripepi}, \&
  {van Loon}}]{schmidt2020}
{Schmidt}, T., {Cioni}, M.-R.~L., {Niederhofer}, F., {et~al.} 2020, \aap, 641,
  A134

\bibitem[{{Skrutskie} {et~al.}(2006){Skrutskie}, {Cutri}, {Stiening},
  {Weinberg}, {Schneider}, {Carpenter}, {Beichman}, {Capps}, {Chester},
  {Elias}, {Huchra}, {Liebert}, {Lonsdale}, {Monet}, {Price}, {Seitzer},
  {Jarrett}, {Kirkpatrick}, {Gizis}, {Howard}, {Evans}, {Fowler}, {Fullmer},
  {Hurt}, {Light}, {Kopan}, {Marsh}, {McCallon}, {Tam}, {Van Dyk}, \&
  {Wheelock}}]{skrutskie2006}
{Skrutskie}, M.~F., {Cutri}, R.~M., {Stiening}, R., {et~al.} 2006, \aj, 131,
  1163

\bibitem[{{Stanimirovi{\'c}} {et~al.}(2004){Stanimirovi{\'c}},
  {Staveley-Smith}, \& {Jones}}]{stanimirovic2004}
{Stanimirovi{\'c}}, S., {Staveley-Smith}, L., \& {Jones}, P.~A. 2004, \apj,
  604, 176

\bibitem[{{Sutherland} {et~al.}(2015){Sutherland}, {Emerson}, {Dalton},
  {Atad-Ettedgui}, {Beard}, {Bennett}, {Bezawada}, {Born}, {Caldwell}, {Clark},
  {Craig}, {Henry}, {Jeffers}, {Little}, {McPherson}, {Murray}, {Stewart},
  {Stobie}, {Terrett}, {Ward}, {Whalley}, \& {Woodhouse}}]{Sutherland2015}
{Sutherland}, W., {Emerson}, J., {Dalton}, G., {et~al.} 2015, \aap, 575, A25

\bibitem[{{Tody}(1993)}]{Tody1993}
{Tody}, D. 1993, in Astronomical Society of the Pacific Conference Series,
  Vol.~52, Astronomical Data Analysis Software and Systems II, ed. R.~J.
  {Hanisch}, R.~J.~V. {Brissenden}, \& J.~{Barnes}, 173

\bibitem[{{van der Marel} {et~al.}(2002){van der Marel}, {Alves}, {Hardy}, \&
  {Suntzeff}}]{vanderMarel2002}
{van der Marel}, R.~P., {Alves}, D.~R., {Hardy}, E., \& {Suntzeff}, N.~B. 2002,
  \aj, 124, 2639

\bibitem[{{van der Marel} \& {Kallivayalil}(2014)}]{vandermarel2014}
{van der Marel}, R.~P. \& {Kallivayalil}, N. 2014, \apj, 781, 121

\bibitem[{Vapnik(2000)}]{Vapnik1995}
Vapnik, V. 2000, The Nature of Statistical Learning Theory, Vol.~8, 1--15

\bibitem[{{Vasiliev}(2018)}]{Vasiliev2018}
{Vasiliev}, E. 2018, \mnras, 481, L100

\bibitem[{{Zivick} {et~al.}(2019){Zivick}, {Kallivayalil}, {Besla}, {Sohn},
  {van der Marel}, {del Pino}, {Linden}, {Fritz}, \& {Anderson}}]{zivick2019}
{Zivick}, P., {Kallivayalil}, N., {Besla}, G., {et~al.} 2019, \apj, 874, 78

\bibitem[{{Zivick} {et~al.}(2021){Zivick}, {Kallivayalil}, \& {van der
  Marel}}]{zivick2020}
{Zivick}, P., {Kallivayalil}, N., \& {van der Marel}, R.~P. 2021, \apj, 910, 36

\end{thebibliography}

\begin{appendix}
\section{Solar reflex motion}
\begin{figure}
   \centering
   \includegraphics[width=\hsize]{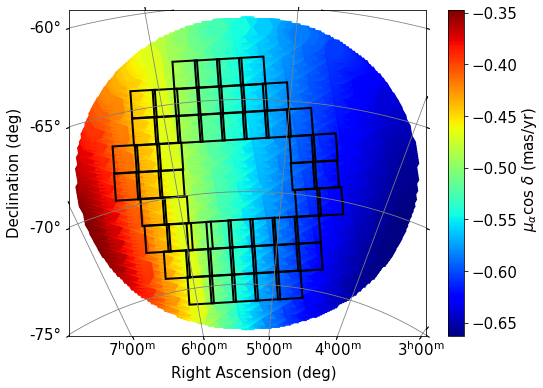}
   \includegraphics[width=\hsize]{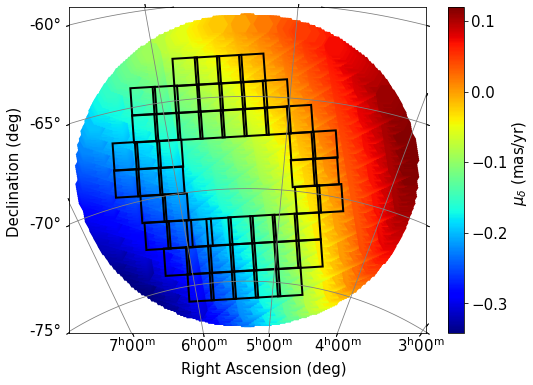}
      \caption{Effect of the solar reflex motion across the LMC on the proper motion components: $\mu_\alpha$cos$(\delta$) (top) and $\mu_\delta$ (bottom) as used in the \textit{Gaia} EDR3 mock catalogue.}
         \label{reflex_motion_vsun_model}
   \end{figure}

\begin{figure*}
   \centering
   \begin{tabular}{cc}
   \includegraphics[width=0.49\textwidth]{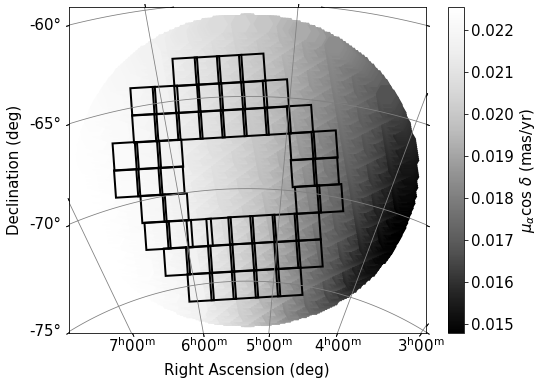}&
   \includegraphics[width=0.49\textwidth]{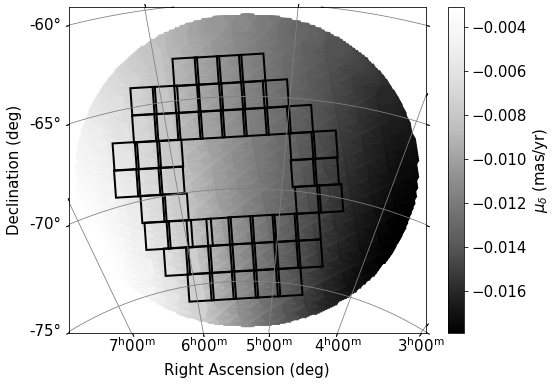}\\
   \end{tabular}
   \caption{Difference between the effect of the solar reflex motion in the \textbf{Gaia} EDR3 mock catalogue and as measured with the data across the LMC on the proper motion components: $\mu_\alpha$cos$(\delta$) (left) and $\mu_\delta$ (right).}
   \label{reflex_motion}
    
    \end{figure*}

The outer regions of the LMC cover large parts of sky. On this scale there is an effect on the proper motions measurements caused by the reflex motion of the sun. The effect is illustrated in Fig.~\ref{reflex_motion_vsun_model}. The difference on the opposite side (west and east) is on the same order as the statistical uncertainties. We found no significant difference, when comparing the reflex motion used in the mock catalogue and the updated values (Fig.\ref{reflex_motion}).
%
%

\section{Source classification}
\begin{figure*}
   \centering
   \begin{tabular}{cc}
   \includegraphics[width=0.48\textwidth]{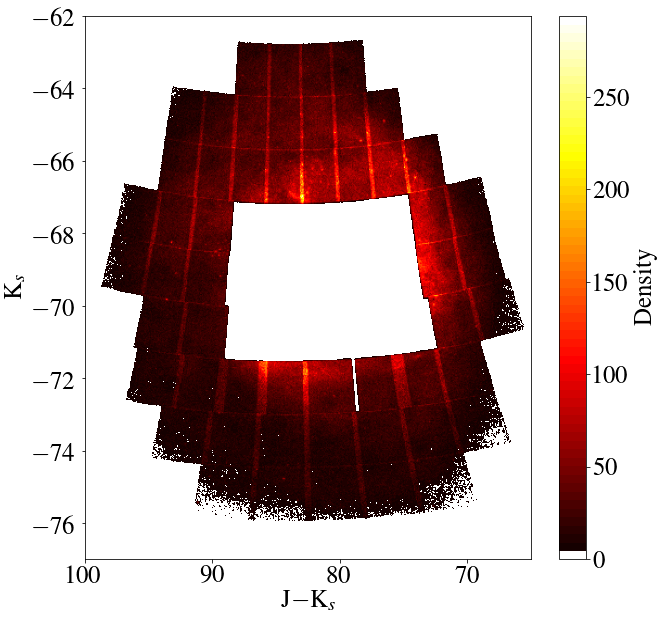}&
   \includegraphics[width=0.48\textwidth]{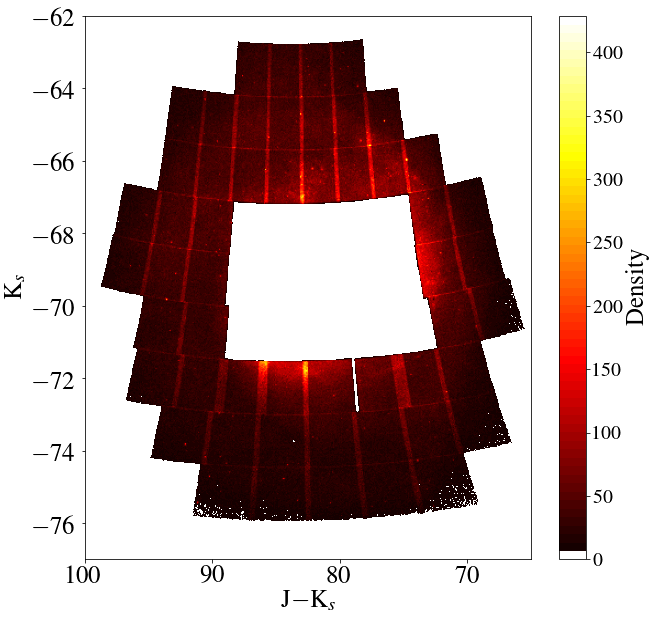}
   \end{tabular}
   \caption{Spatial distribution of sources classified as LMC members (left) and MW foreground stars (right). Higher densities in the overlapping regions of tiles are due to duplication in the VMC data as they represent independent measurements.}
    \label{classification_LMC_map}%
    \end{figure*}

Most LMC sources that were wrongly classified as MW foreground sources (main sequence and RGB stars) can be found in regions with a high stellar density, e.g. close to the bar and the northern spiral arms (see Fig.~\ref{classification_LMC_map}). 
    
   

%

\begin{figure*}
\centering
\begin{tabular}{c c c}
\includegraphics[width=0.3\textwidth]{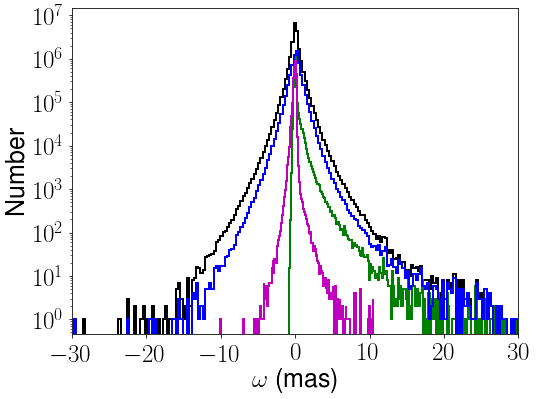}&
\includegraphics[width=0.3\textwidth]{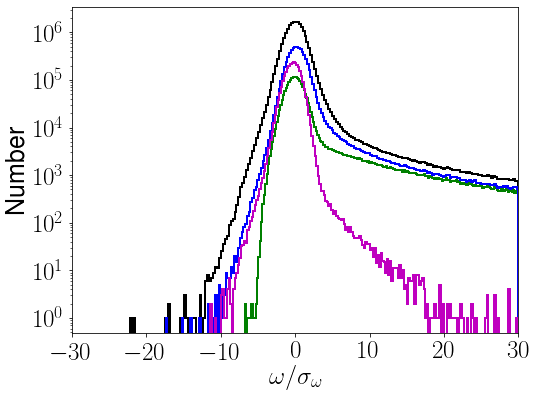}&
\includegraphics[width=0.3\textwidth]{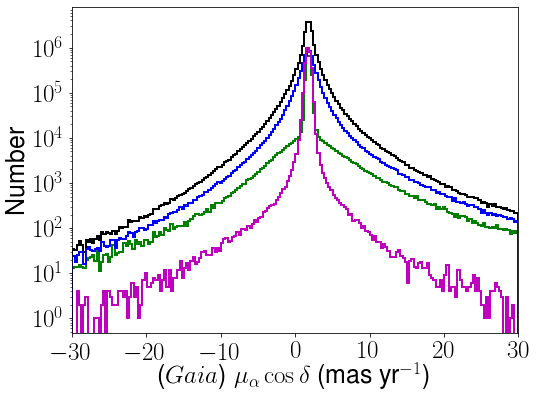}
\\
\includegraphics[width=0.3\textwidth]{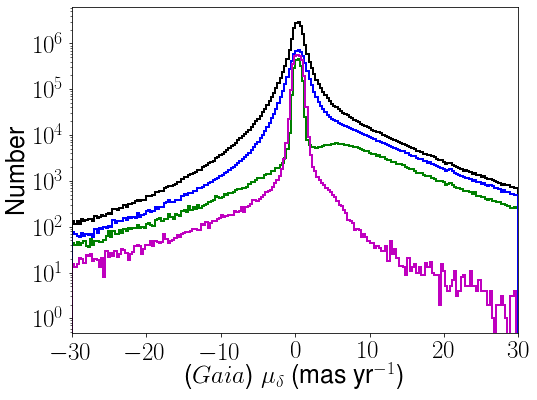}&
\includegraphics[width=0.3\textwidth]{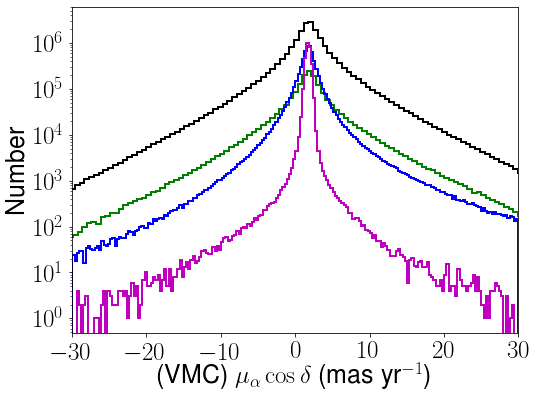}&
\includegraphics[width=0.3\textwidth]{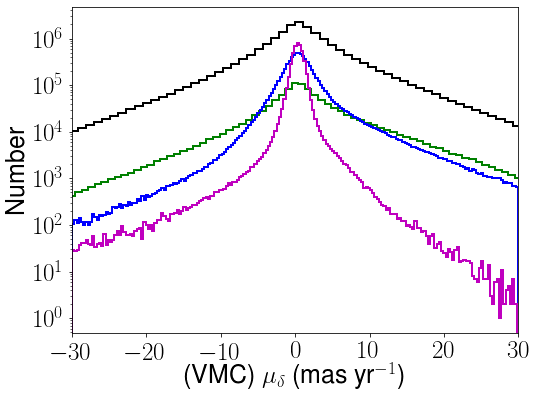}
\\
\includegraphics[width=0.3\textwidth]{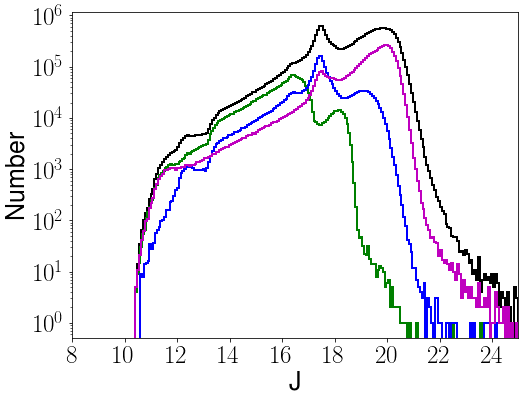}&
\includegraphics[width=0.3\textwidth]{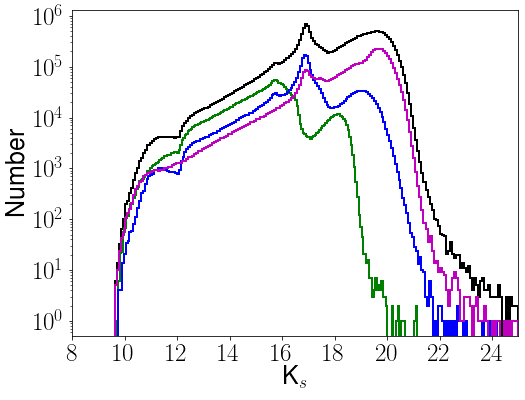}&
\includegraphics[width=0.3\textwidth]{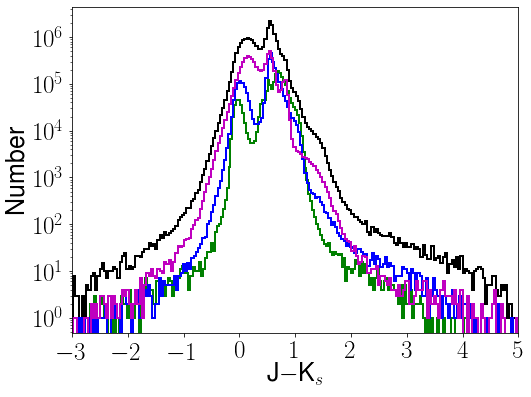}
\\
\includegraphics[width=0.3\textwidth]{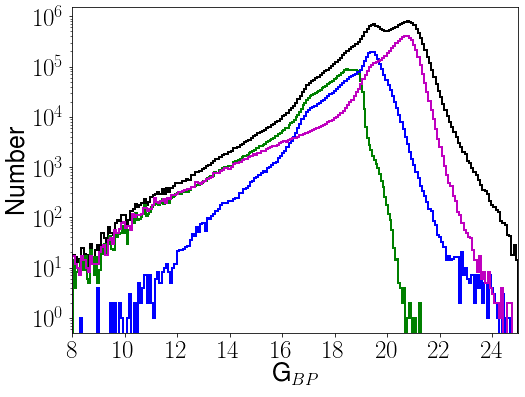}&
\includegraphics[width=0.3\textwidth]{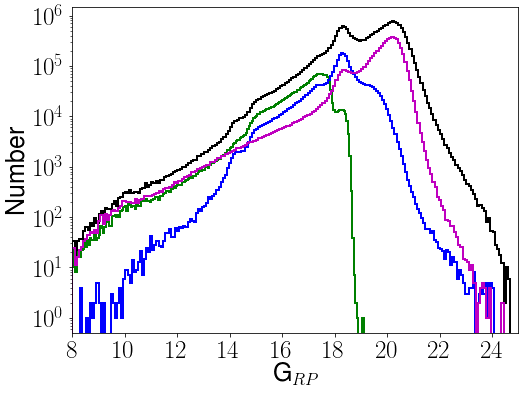}&
\includegraphics[width=0.3\textwidth]{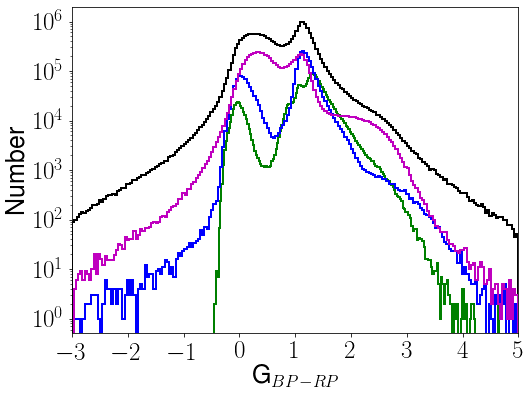}
\\
\includegraphics[width=0.3\textwidth]{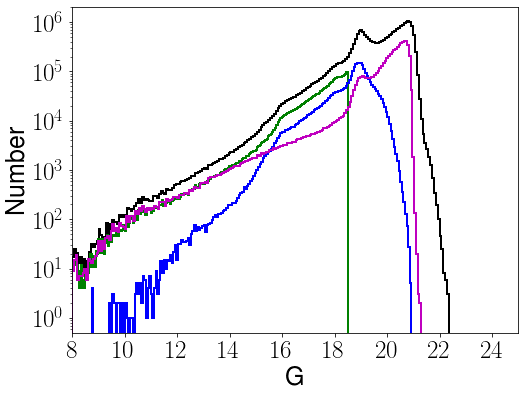}
\end{tabular}
\caption{Histograms of the distribution of the number of stars with respect to the 13 parameters used by the machine learning algorithm. The training sample is shown in green whereas the full catalogue is shown in black. Magenta and blue lines refer instead to the distributions of LMC and MW stars, respectively, as classified by the algorithm.}
\label{MLhistograms}
\end{figure*}

Figure \ref{MLhistograms} compares the distribution of stars from the full catalogue, the training sample and the stars classified by the machine learning algorithm to belong to the LMC and MW, respectively, for each of the 13 parameters used by the algorithm.

\section{VMC and $Gaia$ proper motions}
\begin{figure*}
\centering
\begin{tabular}{cc}
\includegraphics[width=0.45\textwidth]{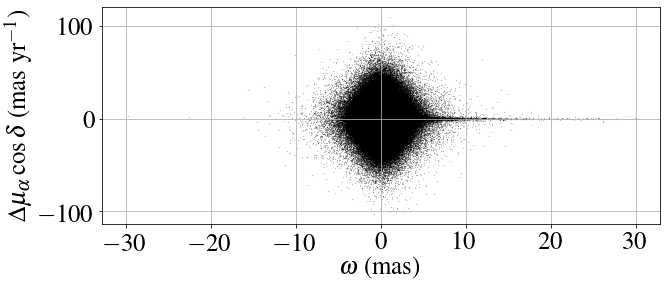}&
\includegraphics[width=0.45\textwidth]{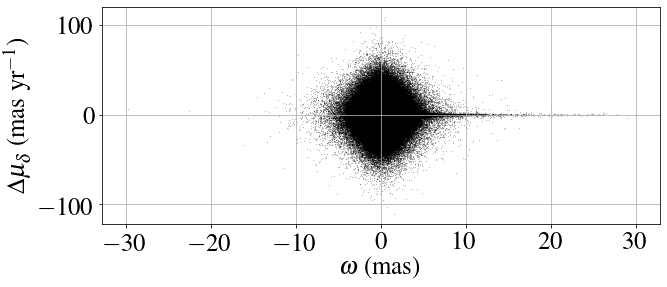}
\\
\includegraphics[width=0.45\textwidth]{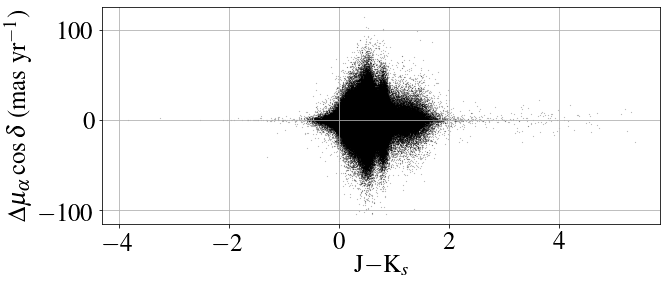}&
\includegraphics[width=0.45\textwidth]{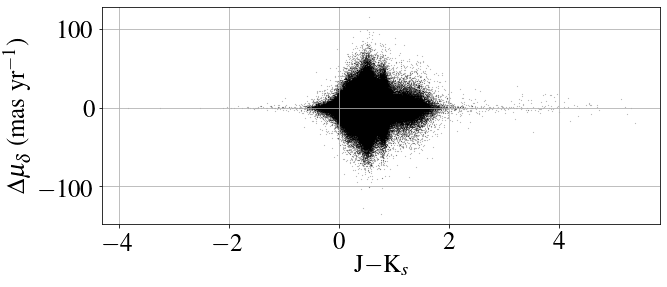}
\\
\includegraphics[width=0.45\textwidth]{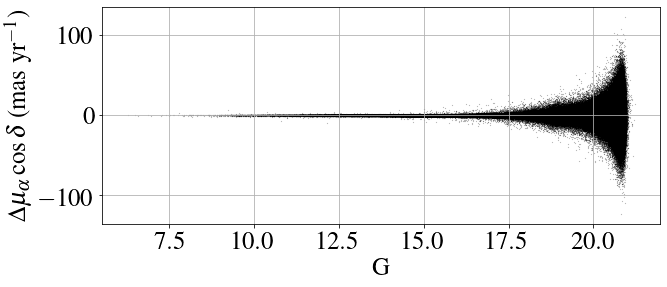}&
\includegraphics[width=0.45\textwidth]{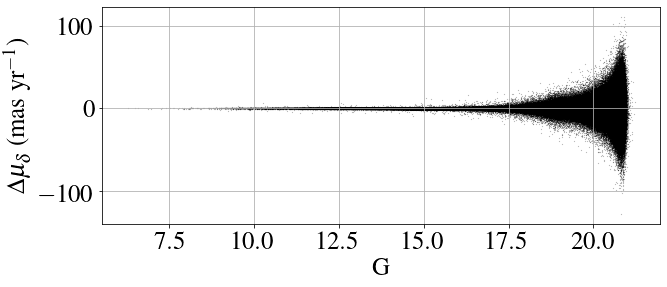}
\\
\includegraphics[width=0.45\textwidth]{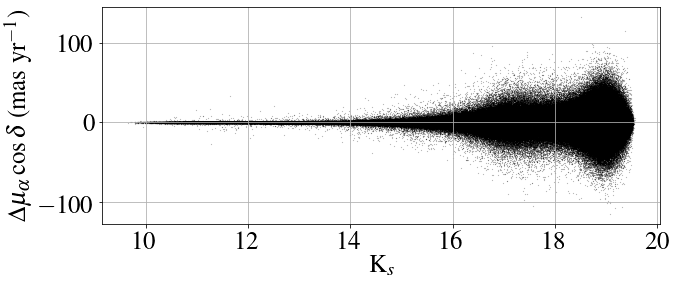}&
\includegraphics[width=0.45\textwidth]{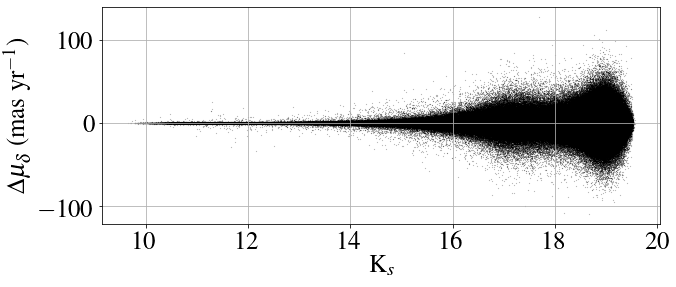}
\\
\includegraphics[width=0.45\textwidth]{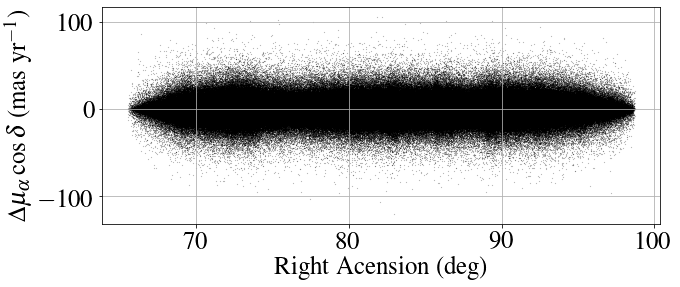}&
\includegraphics[width=0.45\textwidth]{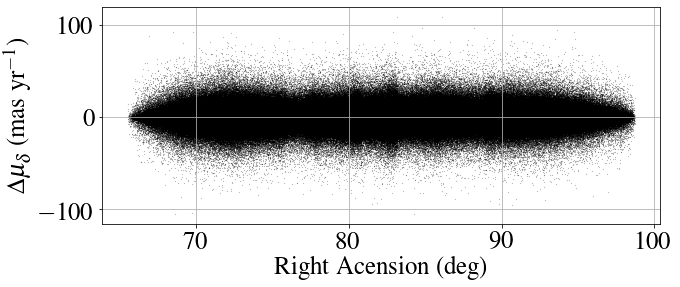}
\\
\includegraphics[width=0.45\textwidth]{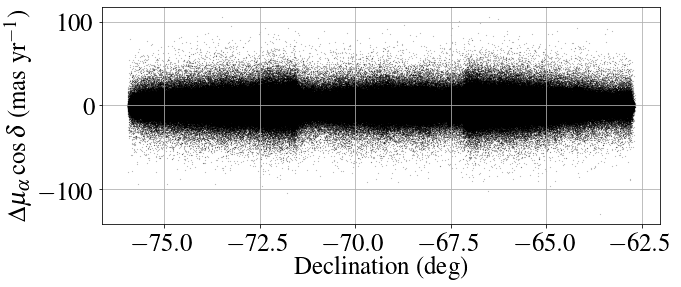}&
\includegraphics[width=0.45\textwidth]{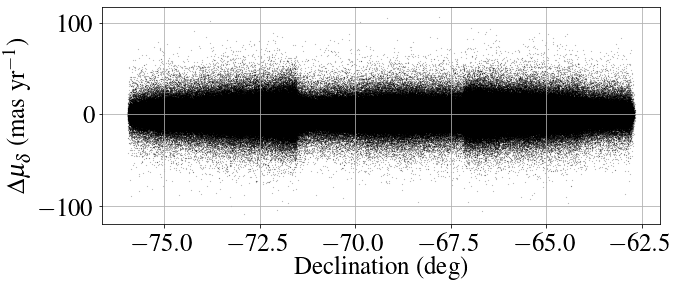}
\end{tabular}
\caption{Proper motion differences between $Gaia$~EDR3 and VMC as a function of: parallax $\omega$, $J-K_\mathrm{s}$ colour, $G$ magnitude, $K_\mathrm{s}$ magnitude, Right Ascension and Declination (from top to bottom) for both components of motion (from left to right).}

\label{comparison_VMC_Gaia}%
\end{figure*}

The proper motion differences between $Gaia$~EDR3 and VMC as a function of parallax, colour, magnitude and position show that there is no obvious correlation (Fig. \ref{comparison_VMC_Gaia}). The consistency between VMC and $Gaia$ proper motions was also discussed in \citet{niederhofer2021}.

\begin{figure*}
   \centering
   \begin{tabular}{cc}
   \includegraphics[width=0.48\textwidth]{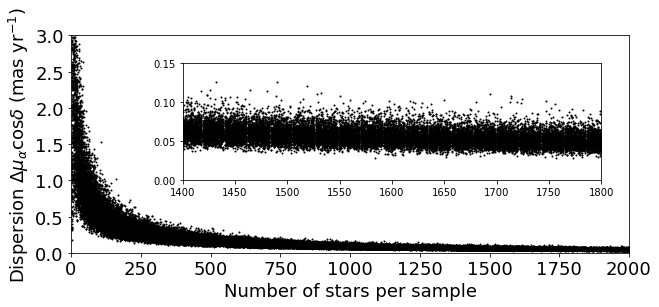}&
   \includegraphics[width=0.48\textwidth]{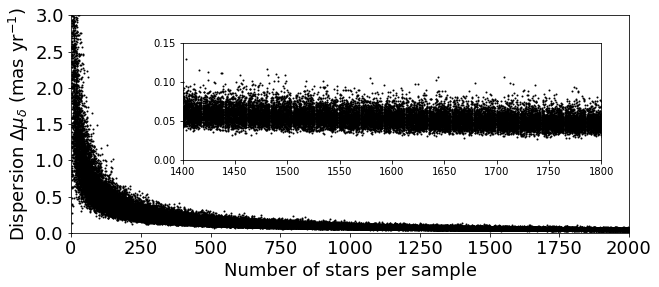}
   \end{tabular}
    \caption{Dispersion of the differences between $Gaia$~EDR3 and VMC proper motion in Right Ascension (left) and Declination (right) for 50 random stellar samples with different sizes as indicated along the horizontal axis. The zoom-in panels are centred at the number of stars corresponding to the smallest bin of Fig.~\ref{LMC_rotaion_VMC}.}
    \label{Dispersion_Gaia_m_VMC}

    \end{figure*}

The dispersion of the difference between VMC and $Gaia$ proper motions in relation to the number of stars in each bin is shown in Fig. \ref{Dispersion_Gaia_m_VMC}. The average dispersion of 50 random samples of stars across the LMC (including both LMC and MW stars) drops below the average measurement uncertainty of individual stars around 100 and approaches zero for roughly 500 stars. Each Voronoi bin in the final VMC catalogues contains significantly more (at least 1600) stars, which indicates a good agreement between the two data sets.

%

\end{appendix}


%
%
\end{document}